\documentclass[12pt,preprint]{aastex}

\begin{document}

\shorttitle{Methods for Estimating Fluxes and Absorptions of Faint X-ray Sources}
\shortauthors{Getman et al.} 
\slugcomment{Accepted for publication in ApJ on 12/01/09}


\title{Methods for Estimating Fluxes and Absorptions of Faint X-ray Sources}

\author{Konstantin V.\ Getman\altaffilmark{1}, Eric D.\ Feigelson\altaffilmark{1,2}, Patrick S.\ Broos\altaffilmark{1}, Leisa K.\ Townsley\altaffilmark{1}, Gordon P.\ Garmire\altaffilmark{1}}

\altaffiltext{1}{Department of Astronomy \& Astrophysics, 525
Davey Laboratory, Pennsylvania State University, University Park
PA 16802}  \altaffiltext{2}{Center for Exoplanets and Habitable Worlds, 525
Davey Laboratory, Pennsylvania State University, University Park
PA 16802}

\email{gkosta@astro.psu.edu}

\begin{abstract}

X-ray sources with very few counts can be identified with low-noise X-ray detectors such as the Advanced CCD Imaging Spectrometer onboard the {\it Chandra X-ray Observatory}. These sources are often too faint for parametric spectral modeling using well-established methods such as fitting with XSPEC. We discuss the estimation of apparent and intrinsic broad-band X-ray fluxes and soft X-ray absorption from gas along the line-of-sight to these sources, using nonparametric methods. Apparent flux is estimated from the ratio of the source count rate to the instrumental effective area averaged over the chosen band. Absorption, intrinsic flux, and errors on these quantities are estimated from comparison of source photometric quantities with those of high signal-to-noise spectra that were simulated using spectral models characteristic of the class of astrophysical sources under study.

The concept of this method is similar to the long-standing use of color-magnitude diagrams in optical and infrared astronomy, with X-ray median energy replacing color index and X-ray source counts replacing magnitude. Our nonparametric method is tested against the apparent spectra of $\sim 2000$ faint sources in the {\it Chandra} observation of the rich young stellar cluster in the  M~17 HII region. We show that the intrinsic X-ray properties can be determined with little bias and reasonable accuracy using these observable photometric quantities without employing often uncertain and time-consuming methods of non-linear parametric spectral modeling.  Our method is calibrated for thermal spectra characteristic of stars in young stellar clusters, but recalibration should be possible for some other classes of faint X-ray sources such as extragalactic active galactic nuclei. 

\end{abstract}


\section{INTRODUCTION \label{introduction_section}}

Many thousands of X-ray sources are discovered using X-ray cameras with modest spectral resolution, such as the CCD detectors on the $ASCA$, $Chandra$, and $XMM$-$Newton$ space observatories.  After detecting and locating the source, the observer often seeks to estimate intrinsic properties of the source, particularly its flux in a broad band of interest and spectral characteristics such as power law index, thermal plasma temperature, or column density of line-of-sight gas absorbing soft X-rays. Low-resolution source and background spectra are extracted from the CCD data, and $\chi^{2}$ or maximum likelihood statistical nonlinear regression of these data with multi-parameter spectral models is performed with codes such as XSPEC\footnote{\url{http://heasarc.gsfc.nasa.gov/docs/xanadu/xspec/}}, Sherpa\footnote{\url{http://cxc.harvard.edu/sherpa/}}, or MIDAS\footnote{\url{http://www.eso.org/sci/data-processing/software/esomidas/}}.  

We have been particularly concerned with the study of large populations of faint X-ray sources associated with pre-main sequence (PMS) stars residing in or near molecular clouds \citep[e.g.][]{Feigelson02,Getman05,Townsley06,Broos07}.  These and other studies of X-ray-bright young stars use a spectral model family involving one- or two-temperature thermal plasmas with a particular pattern of non-solar elemental abundances  \citep[e.g.][]{Maggio07}.  Each star is subject to a different absorption due to its chance location in the molecular cloud and, for the youngest systems, its local protostellar envelope. However, when faint sources are considered, there can be too few counts to obtain reliable spectral fits using regression methods;  a wide range of models can be fit to the same distribution of source photon energies.  While apparent broad-band fluxes can still be estimated, the role of absorption, the intrinsic fluxes, and true plasma temperature(s) are often poorly constrained from spectral modeling of faint sources.  Also, parametric regression analyses are often inhomogeneous in practice; two-temperature plasma models may be fit to lightly-absorbed, bright (say, $>$200 count) sources while fainter and/or highly-absorbed sources may be fit with one-temperature plasma models. There is thus a motivation to consider nonparametric approaches to interpreting X-ray CCD energy distributions which can be applied in a uniform fashion to bright and faint sources.  

In this work, we examine the nonparametric inference of intrinsic X-ray luminosities and absorbing column densities which can be efficiently obtained for large populations of both bright and faint X-ray CCD sources. The concept of the method is similar to the long-standing use of color-magnitude diagrams in optical and infrared astronomy.  For PMS stars, a photometric magnitude and at least one suitable color index, such as the $J$ $vs.$ $J-H$ color magnitude diagram, are used in conjunction with distance, age estimates, and stellar interior evolutionary models to estimate stellar masses, bolometric luminosities, and line-of-sight absorption. In our proposed method, photometric X-ray quantities such as apparent hard-band background-corrected flux (analogous to the $J$ magnitude) and background-corrected total-band median energy (analogous to the $J-H$ color) are used to estimate absorption-corrected X-ray fluxes, and line-of-sight absorbing column densities.  We apply our nonparametric estimates to a large sample of faint X-ray sources in the M~17 star-forming region \citep[][and Townsley et al., in preparation]{Broos07}. Intrinsic spectra for this class of stars are calibrated to the high signal-to-noise $Chandra$ ACIS-I spectra from the sensitive  $Chandra$ Orion Ultradeep Project \citep[COUP;][]{Getman05}.  Reliability of our nonparametric methods, including both systematic bias and statistical uncertainty, are evaluated using simulated X-ray sources with known properties.

The technical concept of avoiding nonlinear parametric modeling, especially in cases of faint sources, is not new. In {\it Chandra} ACIS studies, \citet{Gagne04} used the extracted X-ray mean energy and its comparison with the simulated spectra of one-temperature models to infer some stellar properties. Our method differs in the use of more physically realistic spectral models and the use of the median instead of mean energy to be statistically robust against outliers. \citet{Hong04} also comment that median energies can give effective spectral estimators in X-ray CCD spectroscopy, but they use the median in conjunction with quartiles which can be statistically less reliable for very faint sources.  For example, the 25\%-50\%-75\% energy quartiles are readily estimated for a source with 9 photons (the energies of the $3^{rd}$, $5^{th}$, and $7^{th}$ photons), but are not readily estimated if 8 or 10 photons are present.

This study is closely linked to the {\it ACIS Extract}\footnote{\url{http://www.astro.psu.edu/xray/docs/TARA/ae\_users\_guide.html}} software package for {\it Chandra} ACIS data analysis which is described by \citet{Broos02,Broos09}.  We use {\it ACIS Extract} version 2009-01-27 in the analysis here.

\section {Methods \label{method_section}}

\subsection{Spectral Models for PMS Stars \label{model_section}}

An initial step requires the definition of the intrinsic spectral shape of the population of X-ray sources under study; this may be flux-dependent.  This shape may be obtained from previous spectral analysis of brighter sources of the same class, or from astrophysical models.  In our application to young stars, we use flux-dependent thermal plasma models.  In other applications, such as active galactic nuclei in extragalactic surveys or X-ray binaries in nearby galaxy studies, different spectral shapes would be used (\S \ref{other_classes_section}).
 
We use the COUP spectral results as calibrators of ACIS observations of star populations in more distant star formation regions. Spectral templates derived from bright COUP sources  will be used to translate apparent photometric properties of weak PMS stars into their intrinsic properties. In \S \ref{apparent_aefluxes_section} and \S \ref{mede_section}, we will concentrate on two observable properties of each source: the apparent hard-band X-ray flux, $F_h$, and the X-ray total-band median energy of background-corrected extracted counts, $MedE_t$. Here the hard-band covers the $2.0-8.0$~keV energy range and the total-band covers the $0.5-8.0$~keV energy range. $MedE_t$ and a nonparametric estimator of $F_h$ are automatically calculated from the extracted source photon events by the {\it ACIS Extract} package \citep{Broos02,Broos09}. Later in the study, we consider total-band fluxes and absorption-corrected fluxes.  

We use the COUP spectral fits to establish a parametric spectral model family as a basis for estimating broad-band fluxes of PMS stars.  \citet{Preibisch05} find that two-temperature thermal plasma models provide good fits to high signal-to-noise ACIS data of $\sim 500$ COUP PMS stars.  The temperature of the hot plasma component and emission measure ratio between hot and cool components increase with stellar surface flux, with the temperature of the cool plasma component remaining approximately constant at $kT_1 \sim 0.8$~keV ($T \sim 9$~MK). While it is recognized that the coronae of active stars intrinsically have a distribution of plasma temperatures \citep[e.g.][]{Gudel07}, the two-temperature model is generally adequate for most ACIS-quality spectra. COUP PMS stars follow the general correlation between the temperatures of the hot and cool plasma components seen in main-sequence stars, but with much higher temperatures \citep{Preibisch05}. The 0.8~keV cool plasma component likely describes a persistent coronal structure of PMS stars in the form of compact loops with high plasma density, while the hot plasma component reflects the contribution of PMS flaring activity \citep[][and references therein]{Preibisch05}.

\begin{figure}
\centering
\includegraphics[angle=0.,width=6.8in]{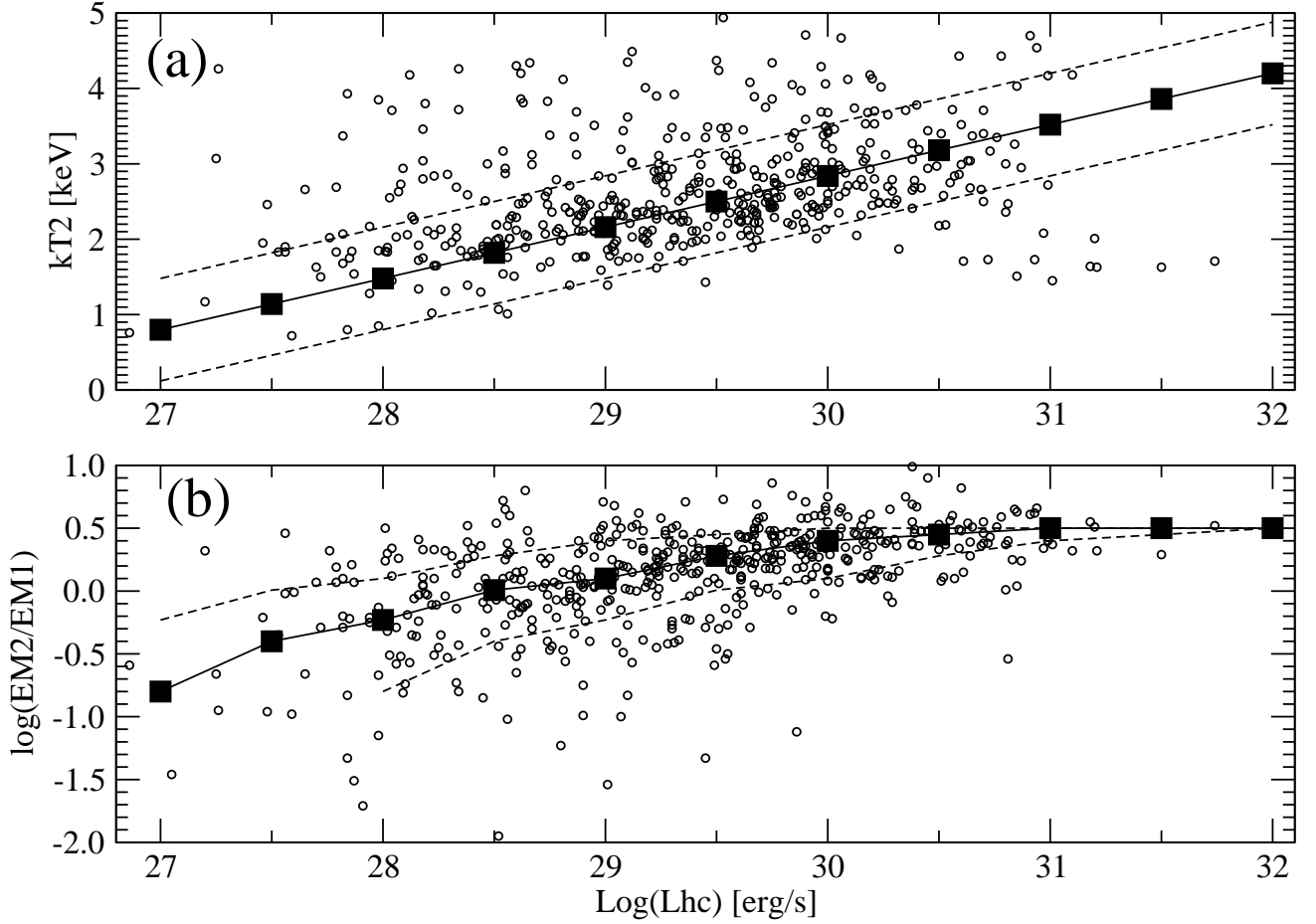}
\caption{Spectral input models (squares) for our simulations based on the COUP data (circles) of $\sim 500$ Orion Nebula Cluster (ONC) PMS stars with available 2-temperature model fits from \citet{Getman05}.  The hot temperature component (panel a) and ratio of emission measures between hot and cool components (panel b) are plotted against intrinsic hard-band ($2.0-8.0$~keV) X-ray luminosity. The dashed curves encompass $\sim 75$\% of the ONC stars and are constructed such that at each $\log(L_{hc})$ bin their y-axis value is equal to the y-axis value of the ``second nearest neighbor'' simulated (square) model; that is, the top (bottom) dashed curves are versions of the solid curves shifted two $\log(L_{hc})$ bins to the left (right). \label {fig_models}}
\end{figure}

Figure \ref{fig_models} presents the dependencies of the hot plasma component, $kT_2$, and the ratio of hot to cool component emission measures, $EM_2/EM_1$, on the absorption-corrected hard-band luminosity, $L_{hc}$, for COUP PMS stars.  A similar diagram appears in Figure 11 of \citet{Preibisch05}.  A few dozen COUP super-hot stars with their $kT_2$ above 5~keV are omitted \citep{Getman08}. The square symbols in the figure and the parameters in Table~\ref{tbl1} define a set of eleven model families we adopt for the class of PMS stars. These are median values with linear extrapolations at the ends of the $L_{hc}$ range\footnote{The spectral trend is compatible with the $L_X \propto T_{av}^{4.3-6.7}$ relation \citep{Telleschi07} seen in the XMM-Newton Extended Survey of Taurus \citep[XEST;][]{Gudel07}, where $T_{av}$ is the mean electron temperature weighted by the differential emission measure distribution and $L_X$ is the intrinsic X-ray luminosity derived in the $0.3-10$~keV energy band.}. The dashed lines around the median values encompass $\sim 75\%$ of COUP stars and indicate the uncertainty in choosing an X-ray model among the model families we consider. This information will be used below to estimate the sensitivity of derived stellar properties on assumed spectral models.  

\begin{figure}
\centering
\includegraphics[angle=0.,width=6.8in]{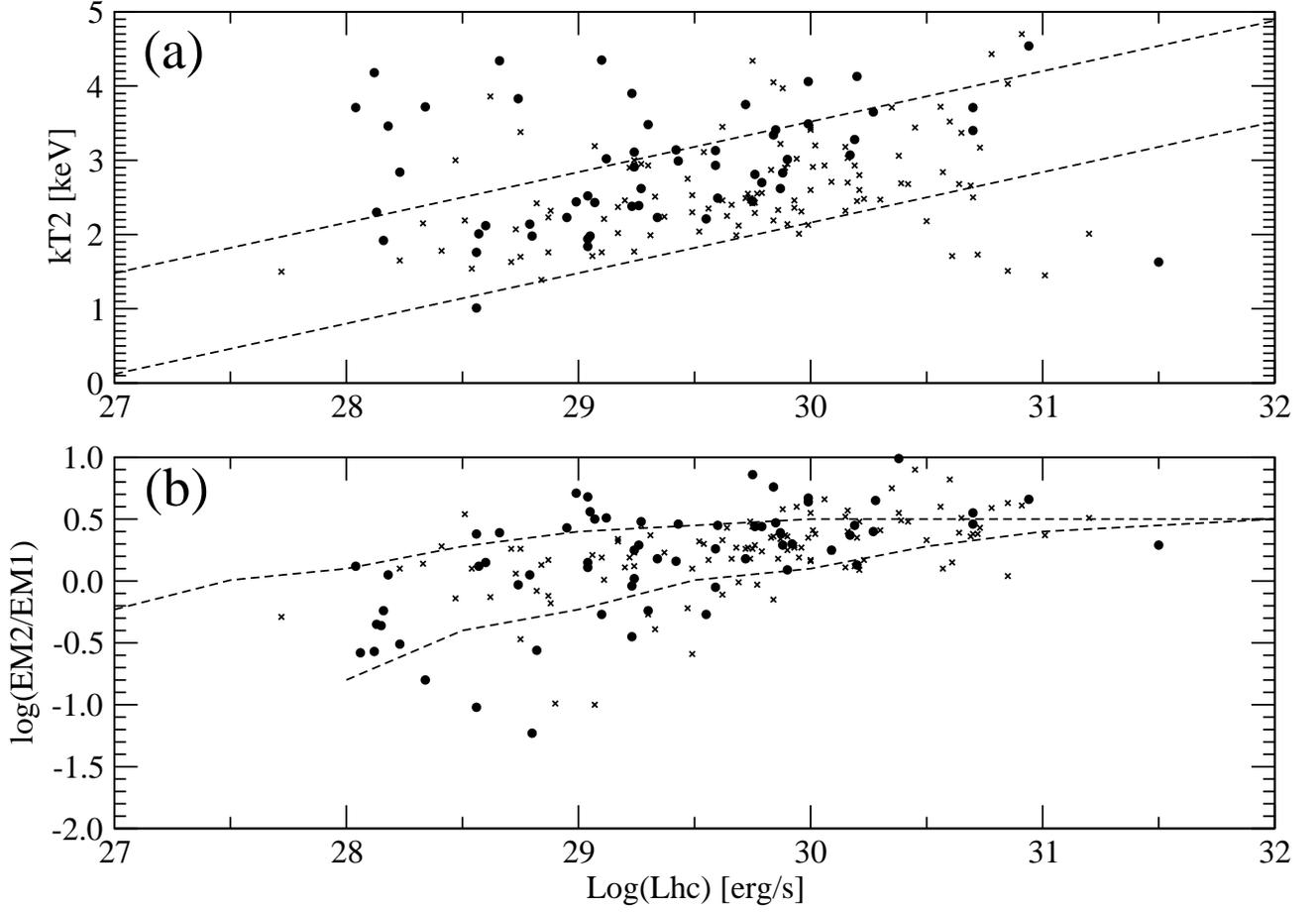}
\caption{Same as Figure \ref{fig_models} but restricted to the COUP PMS star sub-sample with an available accretion indicator \citep[e.g.][]{Preibisch05}. COUP stars classified as active accretors having the Ca~II 8542~$\AA$ line in emission with equivalent width $EW {\rm (Ca~II)} < -1\AA$ are shown with {\large\bf $\bullet$} symbols, and stars classified as weakly accreting or non-accreting with absorption equivalent width $EW \rm{(Ca~II)} > 1\AA$ are shown with $\times$ symbols. \label {fig_models_acc_strata}}
\end{figure}

We note that the XEST project reports little correlation of plasma temperature with X-ray luminosity in Class~II stars (PMS stars with accretion disks), and higher plasma temperatures in Class~II compared to Class~III stars (PMS stars with weak or absent disks) \citep{Telleschi07}.  The COUP sample similarly shows a larger fraction of active accretors lying outside of the band of the X-ray model uncertainty (Figure \ref{fig_models_acc_strata}). Thus the spectral models used here best describe Class~III stars, which generally dominate X-ray-selected samples of young rich X-ray stellar populations. The more general point is that care must be taken to match the spectral model to the underlying population when photometric quantities are examined.

We simulate 11 COUP model families from Table \ref{tbl1} using the {\it fakeit} command in {\it XSPEC} \citep{Arnaud96}. Each model family is characterized by values of $kT_2$ and $EM_2/EM_1$ with $kT_1 = 0.8$~keV and is simulated using the WABS $\times$ (MEKAL$+$MEKAL) {\it XSPEC} model family with 0.3 solar elemental abundances characteristic of young stars\footnote{This simulation was chosen to be compatible with the COUP study \citep{Getman05} where bright lightly absorbed sources were modeled with two-temperature optically thin thermal plasma MEKAL models \citep{Mewe91}, assuming a uniform density plasma with 0.3 times solar elemental abundances \citep{Imanishi01,Feigelson02}. Solar abundances were taken from \citet{Anders89}. X-ray absorption was modeled using the WABS model of atomic cross sections of \citet{Morrison83}.}. For each model family, we generate a large number of high signal-to-noise $Chandra$ ACIS-I spectra over a grid of absorption column densities within the range of $20 \leq \log(N_H) \leq 24$~cm$^{-2}$. Simulated spectra are then passed through the {\it ACIS Extract} code for photometric analysis including calculation of source net counts, median energies, and apparent fluxes in several energy bands.

\subsection{Apparent Photometric Fluxes from {\it ACIS Extract} \label{apparent_aefluxes_section}}

The {\it ACIS Extract} (AE) package has been used in dozens of $Chandra$-ACIS studies, particularly projects involving analysis of many faint sources \citep{Broos09}.  It computes several photometric quantities of interest here from the extracted source events. $NC_t$  and $NC_h$ (in counts) are the total-band and hard-band net counts after subtraction of local background.  $MedE_t$ and $MedE_h$ (in keV) are the total-band and hard-band apparent X-ray median energies after background subtraction. The program estimates photometric photon fluxes (in photons~cm$^{-2}$~s$^{-1}$) in the total and hard bands using
\begin{eqnarray}
F_{t,phot} &=& NC_t/<ARF_t>/Exp \\
F_{h,phot} &=& NC_h / <ARF_h> / Exp. \label{eqn.Fphot}
\end{eqnarray}
Here $<ARF>$ is the mean value of a source's Auxiliary Response File (ARF), the effective area averaged over the appropriate band, and $Exp$ is the exposure time at the source location on the detector\footnote{In the {\it ACIS Extract} documentation, $F_{t,phot}$ is designated $F_2$, while $F_1$ represents another estimator of incident flux where effective areas are measured in narrow energy bands.  The $F_1$-type fluxes are preferred for bright sources but are less stable at low count rates \citep{Broos02,Broos09}. $F_1$-type fluxes are used in the {\it Chandra} Source Catalog where they are called `Aperture Source Energy Fluxes' (\url{http://asc.harvard.edu/csc/columns/fluxes.html}).}.  We estimate the apparent energy fluxes $F_t$ and $F_h$ (in ergs~cm$^{-2}$~s$^{-1}$) as 
\begin{eqnarray}
F_t &=& 1.602 \times 10^{-9} \times F_{t,phot} \times MedE_t \\
F_h &=& 1.602 \times 10^{-9} \times F_{h,phot} \times MedE_h.  \label{eqn.F}
\end{eqnarray}
where $MedE$ are in keV and the constant comes from the keV to ergs conversion. Luminosities $L_t$ and $L_h$ are derived from these fluxes by multiplying them by $4 \pi D^2$ where $D$ is the object's distance, in cm.

In various analyses below, we will consider fluxes that have been corrected for soft X-ray absorption scaled to the line-of-sight column density $N_H$ which we derive from $MedE_t$. These intrinsic absorption-corrected flux estimates are labeled $F_{tc}$ and $F_{hc}$ (in ergs~cm$^{-2}$~s$^{-1}$) for the total and hard bands, respectively.  

\begin{figure}
\centering
\includegraphics[angle=0.,width=6.8in]{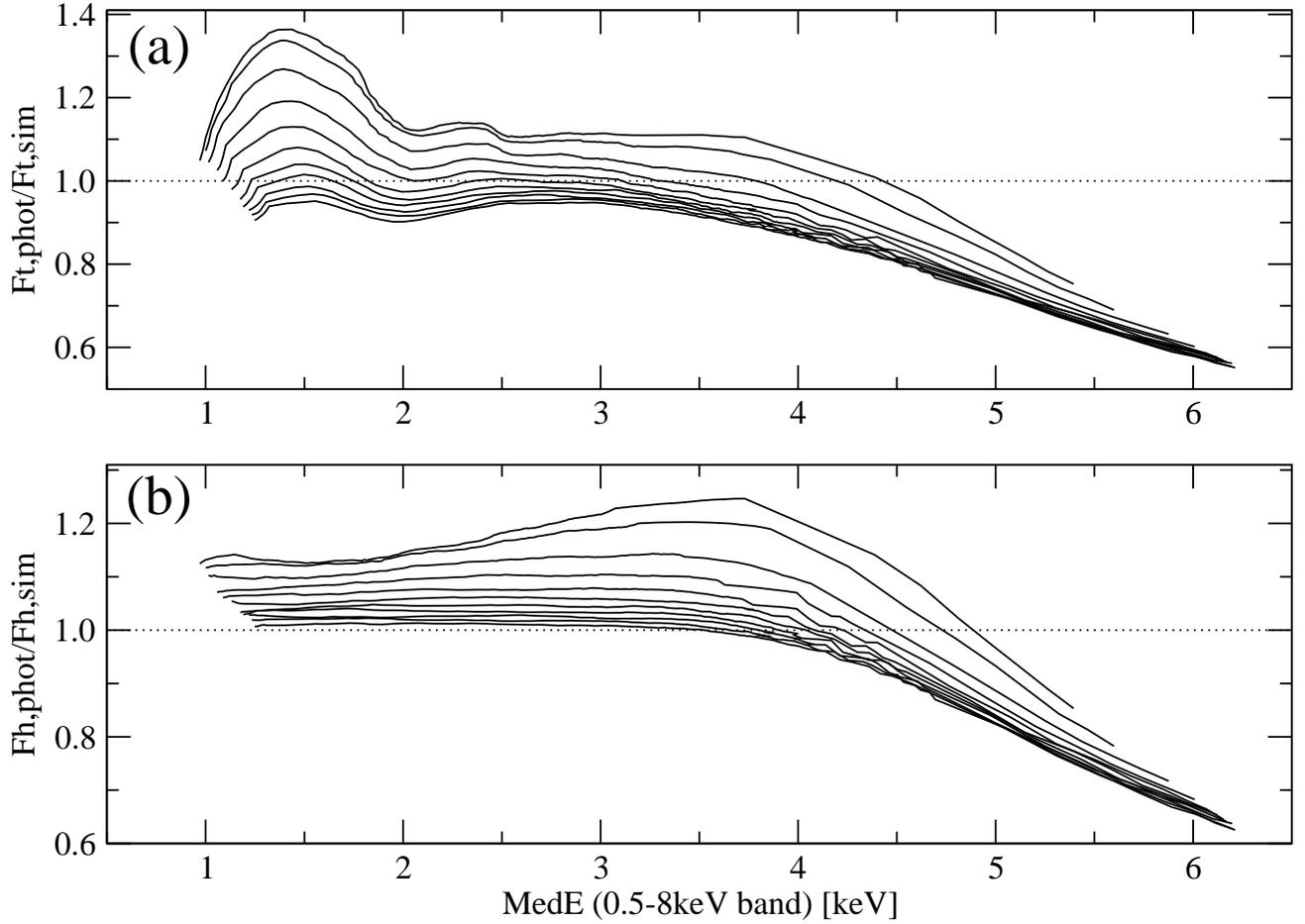}
\caption{Ratio of the apparent X-ray flux of simulated data from {\it ACIS Extract} photometry to that inferred from flux integration of the input spectral models plotted against X-ray median energy in the (a) total band ($0.5-8.0$~keV) and (b) hard band ($2.0-8.0$~keV). Results are presented for all 11 simulated spectral model families with the model families for $\log L_{hc} =27$~ergs~$s^{-1}$ ($\log L_{hc} =32$~ergs~$s^{-1}$) stars as the top (bottom) curves. \label{fig_xspec_vs_phot_sim}}
\end{figure}

In oder to assess the fidelity of these flux estimates, Figure \ref{fig_xspec_vs_phot_sim} compares {\it ACIS Extract} photometric fluxes from equations (3)-(4) for simulated data with apparent fluxes derived from the flux integration of the input models (discussed in \S\ref{model_section}) evaluated on the energy bins set by the {\it Chandra} response matrix. Discrepancies of $10-20$\% in flux are seen for typical spectra, and can range up to $\sim 40$\% for extremely soft or hard spectra. These biases can be explained by inaccuracies in the computation of fluxes using equations (1)-(2) that are based on effective areas averaged over the broad band assuming a flat incident spectrum.  This approximation is not accurate for realistic spectral shapes.  {\it ACIS Extract} $F_1$-type incident fluxes, where effective areas are measured in narrow energy bands, do not exhibit these biases, but they suffer from large Poisson errors for weak sources. For example, a single photon present at energies $E \ga 6$ keV, where the {\it Chandra} mirror response is poor, can lead to a large spurious jump in inferred flux. Since our work is oriented towards study of faint sources, we use the more stable $F_2$-type photometric fluxes of equations (1)-(2). The curves in Figure \ref{fig_xspec_vs_phot_sim}  can then be used to reduce the systematic biases in flux extimates for extremely soft and hard spectra (see also Table~\ref{tbl_calibration1}).

\subsection{Association Between Apparent Median Energy and Inferred Absorption \label{mede_section}}

$F_t$ and $F_h$ are not reliable estimators of the intrinsic source flux if absorption is present.  For bright sources, a typical procedure is to estimate line-of-sight column density $N_H$ via a parametric fit of the apparent photon energy distribution using {\it XSPEC} or similar code, and calculate an intrinsic unabsorbed flux based on the other spectral model parameters.  We now seek an analogous procedure based on the apparent source median energy, $MedE_t$.    

Figure \ref{fig_nh_vs_mede_sim} (see also Table~\ref{tbl_calibration2}) shows the simulated X-ray column density ($N_H$) as a function of the X-ray source's median energy in the total band ($MedE_t$). The results of our simulations confirm the previous COUP finding of \citet{Feigelson05} that the median energy can be effectively used as a surrogate for absorption column density for $MedE_t \ga 1.5-1.7$~keV corresponding to $\log N_H \ga 22.0$ cm$^{-2}$.  The 0.5~keV low energy limit of the ACIS instrument leads to a degeneracy in absorption values for $MedE_t \sim 1.0-1.3$~keV.   The $MedE_t-\log N_H$ relationship can be used reliably to infer either absorbing column density or source plasma temperature if $N_H$ is known in advance.  If both are unknown, the inference has greater error (\S \ref{method_error_section}).  

Figure \ref{fig_redcorr_vs_mede_sim} (see also Table~\ref{tbl_calibration3}) shows the ratio of intrinsic (absorption-corrected) flux to apparent flux against apparent median energy in the total band for the simulated spectra. These are flux integrations of the input models (not re-fitted simulated spectra) evaluated on the energy bins set by the {\it Chandra} response matrix. We see that flux ratios for the total band fluxes,  $F_{tc}/F_t$,  are often factors of $\sim 10$, and can reach factors of $\sim 30$, for heavily absorbed sources.  In the hard band, the flux ratios $F_{hc}/F_h$ are much smaller, often less than a factor of 2 and always less than a factor of 10. $F_{hc}$ is thus a much more stable estimate of source flux than $F_{tc}$ when absorption is present.

\subsection{Implementation \label{implementation_section}}

As described above, the inferred relation between the intrinsic spectral shape and the X-ray luminosity (\S \ref{model_section} and Figures \ref{fig_models} and \ref{fig_models_acc_strata}) allowed us to explicitly define and simulate X-ray models of PMS stars. Analysis of the simulated data calibrates relationships between spectral models and observable quantities (\S \ref{apparent_aefluxes_section} and \ref{mede_section}, and Figures \ref{fig_xspec_vs_phot_sim}, \ref{fig_nh_vs_mede_sim}, \ref{fig_redcorr_vs_mede_sim}). We have shown that the source counts and median energy, which are readily measured for faint sources, often serve as surrogates for source flux and absorption. Similar observable quantities have been used for many years in color-magnitude diagrams for characterizing sources in optical-infrared (OIR) astronomy. 

Other classes of sources, different from young stars, may have their own relationships between intrinsic and/or observable quantities that allow simulations of explicit X-ray models (\S \ref{other_classes_section}). Once model simulations are performed, calibration data mapping simulated observable to simulated intrinsic properties can be built. In the case of PMS stars we produce four calibration products: Tables \ref{tbl_calibration1}, \ref{tbl_calibration2}, and \ref{tbl_calibration3} correspond to Figures \ref{fig_xspec_vs_phot_sim}-\ref{fig_redcorr_vs_mede_sim} discussed above, and Table \ref{tbl_calibration4} is used for error analysis (\S \ref{method_error_section}).


Once simulated calibration data are built, the measured photometric properties of observed X-ray sources are compared with the calibration data to obtain intrinsic properties and errors on these properties. The procedure is shown in Figures \ref{fig_flow_chart_part1} and \ref{fig_flow_chart_part2} and is presented in detail in this section. The IDL implementation of the procedure can be found at \url{http://www.astro.psu.edu/users/gkosta/XPHOT/}. Researchers working with other classes of sources can use the IDL code after constructing the four calibration tables similar to Tables \ref{tbl_calibration1}-\ref{tbl_calibration4} appropriate for the spectral shapes of their source class.

\subsubsection{Translating Apparent to Intrinsic Properties \label{translate_section}}

In Step 1 of the procedure shown in Figure \ref{fig_flow_chart_part1}, source photometry properties are obtained by running {\it ACIS Extract}. Here we provide details on Steps 2-4 which estimate the source's absorption and intrinsic broad-band fluxes.

In Step 2, one of the 11 simulated PMS spectral model families listed in Table \ref{tbl1} is assigned to each individual source of interest based on that source's hard-band apparent luminosity and full-band median energy. For lightly obscured stars ($MedE_t \la 2$~keV), the ratio of intrinsic to apparent flux in the hard-band is $<0.2$~dex for all spectral models (Figure \ref{fig_redcorr_vs_mede_sim}b), so that apparent hard-band luminosity $L_h$ only slightly underestimates the intrinsic hard-band source luminosity $L_{hc}$. The $<0.2$~dex luminosity difference is much less than the quantization of 0.5~dex on $L_{hc}$ used to create the COUP template models (Figure \ref{fig_models} and Table \ref{tbl1}). Thus, for a lightly obscured PMS star, the apparent luminosity $L_h$ can be used directly to assign a specific spectral model family (``best model family'') from those listed in Table \ref{tbl1}.

For a heavily obscured PMS star ($MedE_t > 2$~keV), the source's median energy $MedE_t$ is compared to the calibration of intrinsic to apparent flux ratios (Figure \ref{fig_redcorr_vs_mede_sim}b  and Table \ref{tbl_calibration3}) to choose 11 candidate $F_{hc}/F_{h}$ values corresponding to 11 template model families. Then, the source's apparent hard-band luminosity $L_h$ is used to obtain 11 candidate intrinsic hard-band luminosities, $L_{hc,candidate}$. The ``best model family'' is the one with the closest match between the $L_{hc,candidate}$ and the ``nominal'' intrinsic hard-band luminosity listed in Table \ref{tbl1}.  This procedure is analogous to the reddening correction in OIR color-magnitude diagrams.

In Step 3, photometric fluxes are de-biased using calibration curves (Figure \ref{fig_xspec_vs_phot_sim} and Table \ref{tbl_calibration1}) for the source's spectral model.  The absorbing column density is then estimated using the source's median energy (Figure \ref{fig_nh_vs_mede_sim} and Table \ref{tbl_calibration2}).  Step 4 uses the resulting $N_H$ estimate to obtain absorption-corrected fluxes (Figure \ref{fig_redcorr_vs_mede_sim} and Table \ref{tbl_calibration3}). 

To visually represent the transformation of apparent to intrinsic stellar properties in our method, Figure \ref{fig_xraycmd} shows the X-ray ``color-magnitude diagram'' derived from the simulated PMS star data. In this diagram, the apparent hard-band luminosity ($L_h$), apparent total-band median energy ($MedE_t$), absorbing column density ($N_H$), and intrinsic hard-band luminosity ($L_{hc}$) are analogues to OIR magnitude, color, extinction, and bolometric luminosity, respectively.  We note again the degeneracy in low values of absorbing column density at $MedE_t \sim 1-1.3$~keV due to the 0.5~keV limit of the {\it Chandra}-ACIS instrument.

\subsubsection{Error Analysis \label{method_error_section}}     

We estimate uncertainties in source properties derived using our photometric procedures including both small-$N$ statistical errors and systematic uncertainties in our knowledge of the X-ray spectral model.  Systematic errors associated with the astronomical source properties, such as distance or X-ray variability, are not considered here.  We also omit systematic errors of instrumental origin.  Errors in the ARF, which is based on calibration of the {\it Chandra} mirrors and ACIS detector, are anticipated to be small and become comparable to statistical uncertainties only for sources with $>10,000$ counts \citep{Drake06}.  The procedures for the error analysis of photometric quantities described here are outlined in Figure \ref{fig_flow_chart_part2}.

First, we estimate statistical errors on the X-ray median energy statistic using simulated data. Passing simulated model spectra (\S \ref{model_section}) through the MARX mirror-detector simulator\footnote{\url{http://space.mit.edu/CXC/MARX/}}, we slice the resulting X-ray event lists into thousands of subsamples of different sizes.  Distributions of the median energy statistic are then obtained and corresponding standard errors, $\Delta MedE$, encompassing 68\% of the values are recorded.  The resulting standard errors for the total-band and hard-band median energies for the 11 PMS spectral model families are shown in Figure~\ref{fig_mede_errors} and tabulated in Table \ref{tbl_calibration4} \footnote{We also checked our error results using a different method which directly employs generated spectra without the need of MARX simulations. Here we generate a large number of simulated spectra with the desired number of counts using random deviates from the known energy distributions for each spectral model \citep{Press92}.  The resulting median energy distributions and standard error values are indistinguishable from those obtained from the MARX simulations.}.   

\begin{figure}
\centering
\includegraphics[angle=0.,width=6.8in]{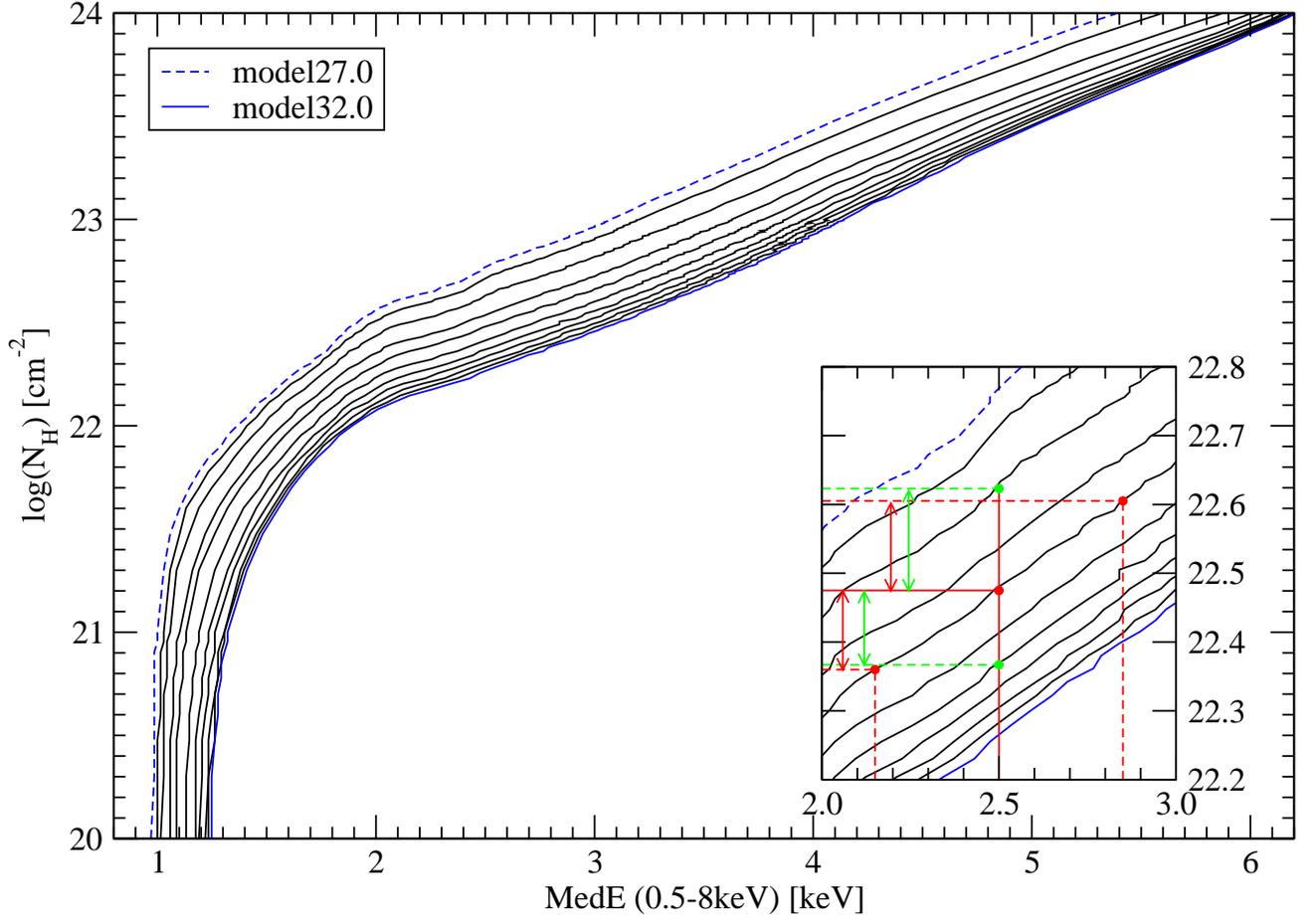}
\caption{Simulated calibration curves of X-ray column density plotted against X-ray median energy, shown for the 11 simulated spectral model families from Figure \ref{fig_models}. The top (blue dashed) and bottom (blue solid) curves correspond to the model families for $\log(L_{hc}) = 27$~ergs s$^{-1}$ and $\log(L_{hc}) = 32$~ergs s$^{-1}$ stars, respectively. The insert exemplifies estimation of the column density, $\log N_{H,phot}$, and its statistical (red arrows) and systematic (green arrows) errors.  In this example, the source has 20 counts with  median energy $2.5$~keV and median energy uncertainty $0.35$~keV, using the $\log(L_{hc}) \sim 29$~ergs s$^{-1}$ spectral model family. \label{fig_nh_vs_mede_sim}}
\end{figure}

\begin{figure}
\centering
\includegraphics[angle=0.,width=6.0in]{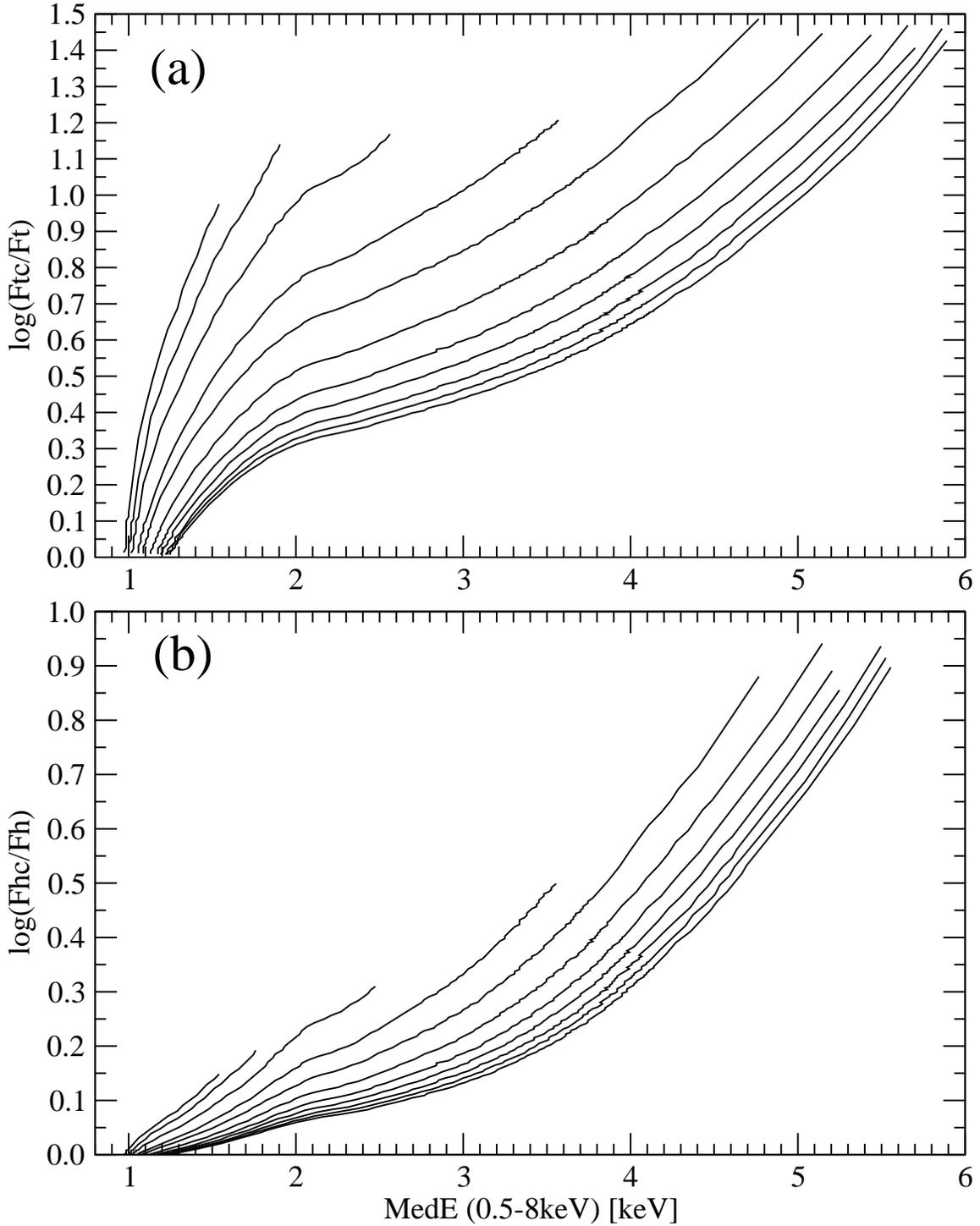}
\caption{Intrinsic to apparent flux ratio of simulated data are plotted against X-ray median energy. Results are shown for the 11 simulated spectral model families (solid lines) with $\log(L_{hc}) = 27$~ergs s$^{-1}$ and $\log(L_{hc}) = 32$~ergs s$^{-1}$ as the leftmost and rightmost curves, respectively. Model curves are truncated to indicate the locus of $\sim 1400$ ONC PMS stars from the sensitive COUP project. Correction factors are shown for the (a) total band ($0.5-8.0$ keV) and (b) hard band ($2.0-8.0$ keV). \label{fig_redcorr_vs_mede_sim}}
\end{figure}

\begin{figure}
\centering
\includegraphics[angle=0.,width=6.8in]{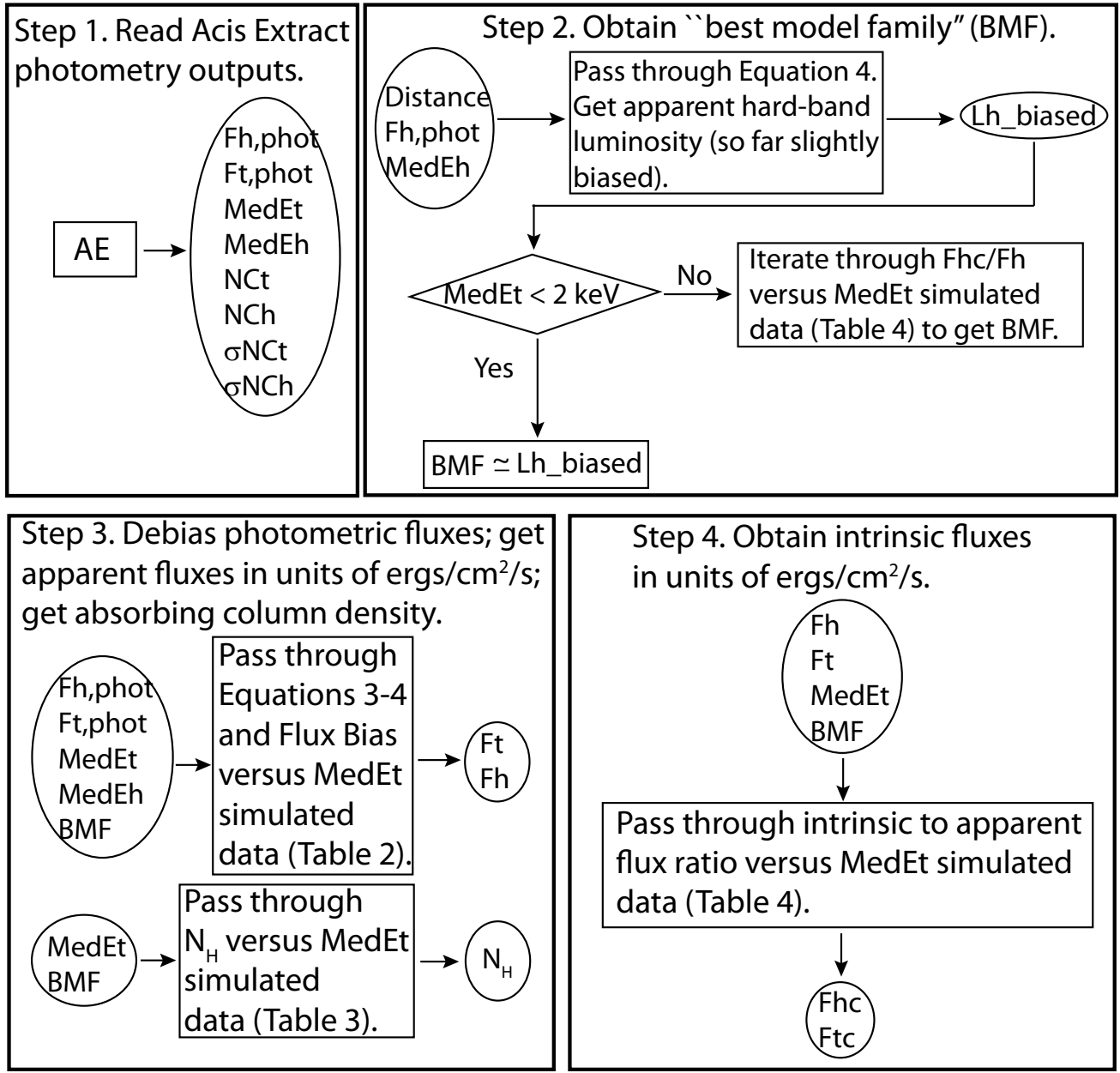}
\caption{The procedure of our nonparametric method given in the form of a flowchart. The procedure for obtaining intrinsic fluxes and absorbing column density. \label{fig_flow_chart_part1}}
\end{figure}

\begin{figure}
\centering
\includegraphics[angle=0.,width=6.8in]{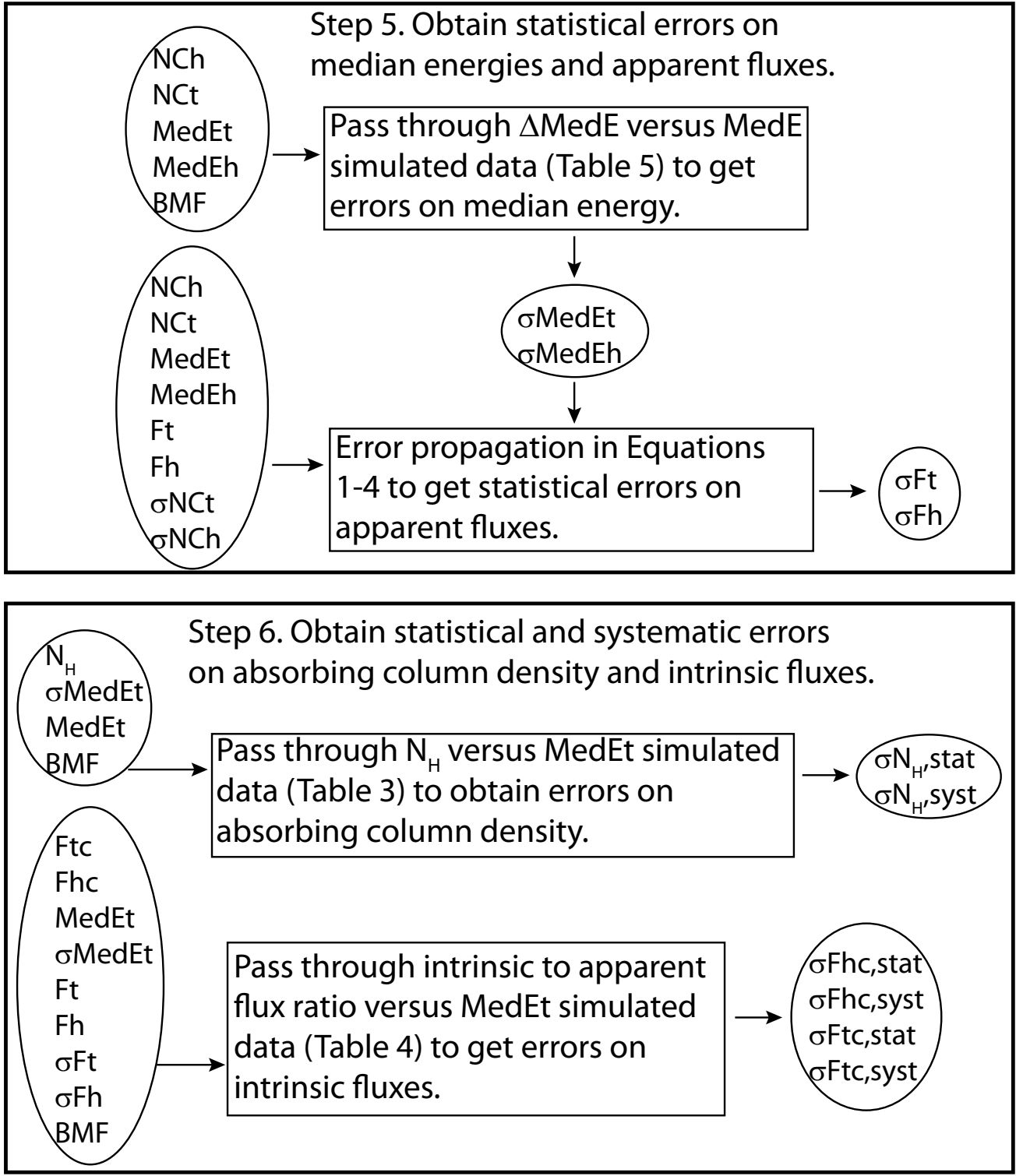}
\caption{The procedure for obtaining errors on intrinsic fluxes and absorbing column density. \label{fig_flow_chart_part2}}
\end{figure}

\begin{figure}
\centering
\includegraphics[angle=0.,width=6.8in]{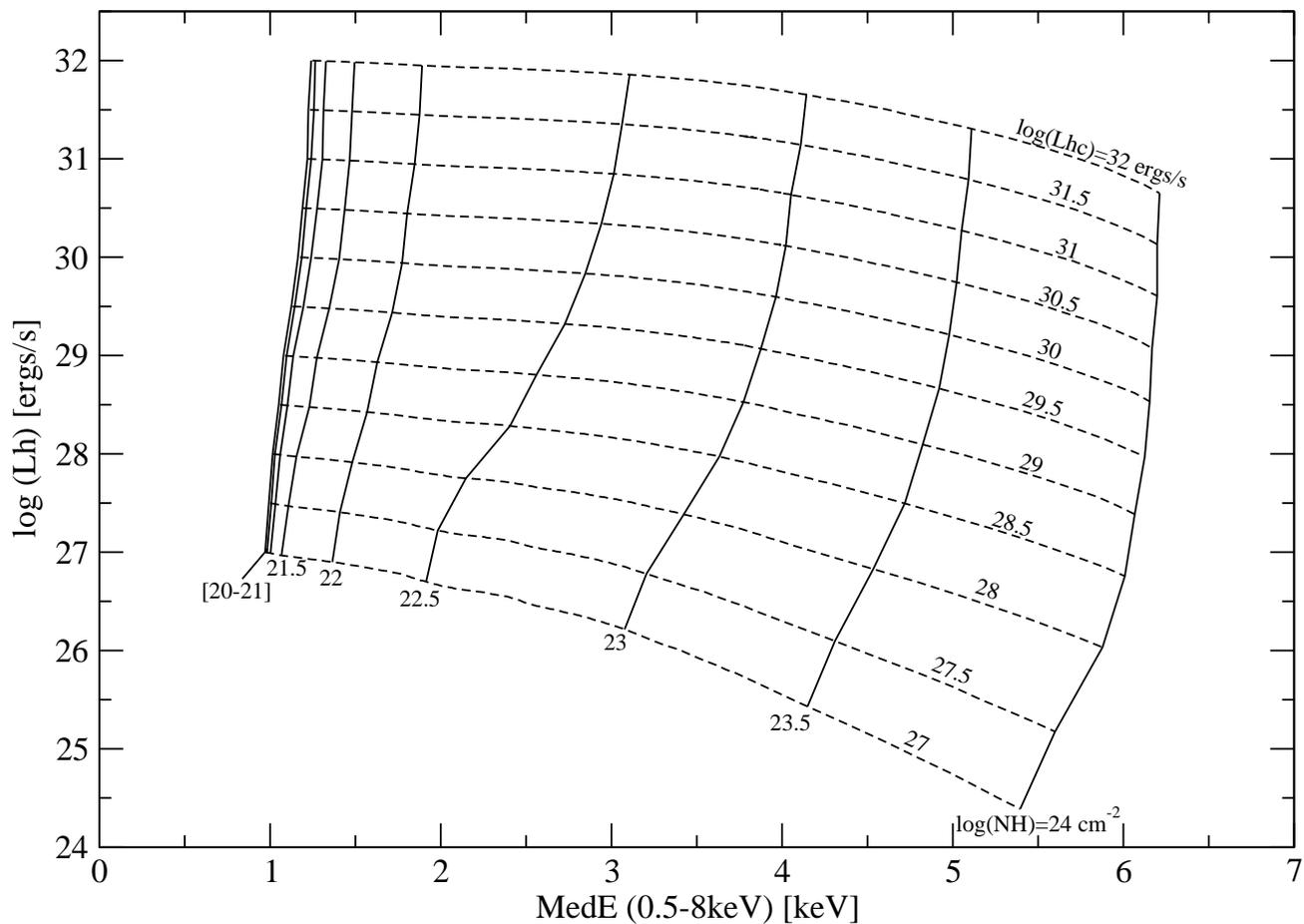}
\caption{X-ray ``color-magnitude diagram'' derived from our method with apparent hard-band flux versus apparent total-band median energy. Lines represent X-ray model isochrones derived from our nonparametric method: dashed lines indicate isochrones of intrinsic hard-band X-ray luminosity, solid lines indicate isochrones of absorbing column density. \label{fig_xraycmd}}
\end{figure}

\begin{figure}
\centering
\includegraphics[angle=0.,width=5.5in]{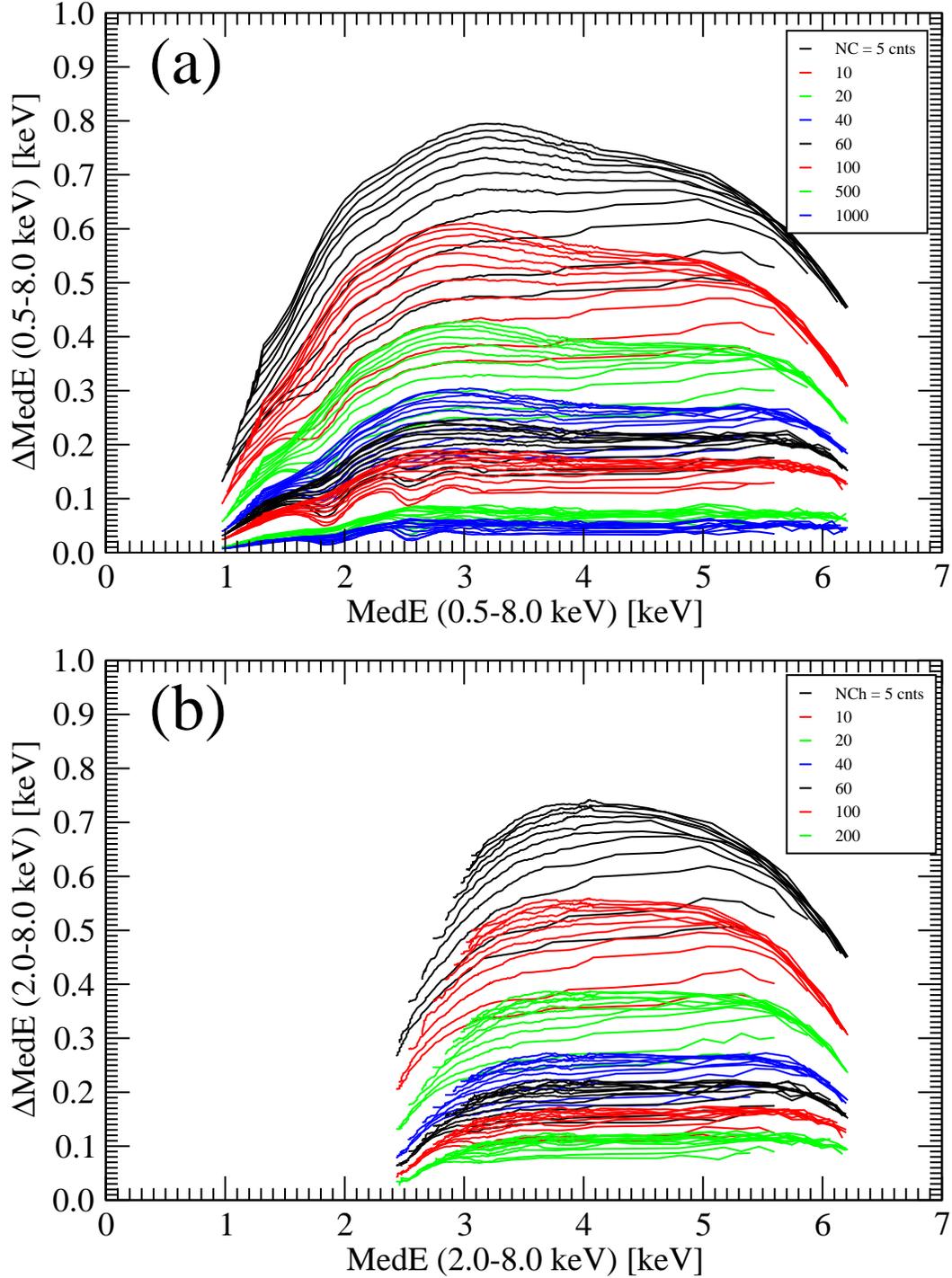}
\caption{Uncertainties of the estimated X-ray median energy plotted against median energy for different net count strata from simulated data in the (a) total band ($0.5-8.0$ keV) and (b) hard band ($2.0-8.0$ keV). For each count stratum, shown with a different color, results are shown for the 11 simulated spectral model families from Figure \ref{fig_models} with $\log(L_{hc}) = 27$~ergs s$^{-1}$ and $\log(L_{hc}) = 32$~ergs s$^{-1}$ as the bottom and top curves, respectively.  Moving along an individual model family curve from left to right corresponds to increasing the column density parameter of that spectral model family (see Figure \ref{fig_nh_vs_mede_sim}). \label{fig_mede_errors}}
\end{figure}

Figure \ref{fig_mede_errors} shows that median energy errors range from $\sim 5$\% for  sources with 100 counts to $\sim 15$\% for sources with 10 counts.  These values are larger than expected from a standard Gaussian distribution, although the errors do scale roughly with $\sqrt{N}$.  At very low ($MedE_t \la 2$~keV) and very high ($MedE_t \ga 5.5$~keV) median energy values, errors are reduced due to the energy limits of the $Chandra$ instrumental response. The spread in $\Delta MedE_t$ for a fixed source count and $MedE_t$ value shows that sources with harder intrinsic spectra have larger median energy uncertainties than intrinsically softer sources\footnote{ When a low-temperature and a high-temperature source have the same observed median energy, the low-temperature source has to be more highly absorbed (Figure \ref{fig_nh_vs_mede_sim}). That low-energy $N_H$ cut-off and the intrinsic lack of power at high energies give the low-temperature source a narrower observed spectrum, and a correspondingly more accurate estimator for apparent median energy.}.

Again we recall that the results here are specific to the {\it Chandra} ACIS mirror-detector combination and PMS stellar spectral models. Researchers working with data from other X-ray telescopes or source classes are advised to produce new error estimates for the median energy statistic. However, we expect that the qualitative findings will be similar for other systems.

To estimate the effects of these median energy uncertainties (Figure \ref{fig_mede_errors} or Table \ref{tbl_calibration4}), we propagate them along with uncertainty on source net counts to errors on apparent fluxes (equations 1-4) using standard methods \citep[e.g.][]{Bevington92}.  This is Step 5 of the procedure shown in Figure \ref{fig_flow_chart_part2}. 

Step 6 of the procedure propagates the median energy errors through the relations between median energy and absorbing column density (Figure \ref{fig_nh_vs_mede_sim} and Table \ref{tbl_calibration2}) to estimate both statistical and systematic errors in $\log N_H$. This is illustrated in the insert diagram of Figure \ref{fig_nh_vs_mede_sim}. Statistical errors on $\log N_H$ are derived by propagating errors on $MedE_t$ using the curve for the ``best model family'' (\S \ref{translate_section}). Systematic errors on $\log N_H$ are estimated by propagating the value of the $MedE_t$ itself to other possible spectral models; we use the ``second nearest neighbor'' as a plausible range for model uncertainty (see the dashed lines in Figure \ref{fig_models} which are discussed in \S \ref{model_section}).

Finally, we estimate statistical and systematic errors on the absorption-corrected hard-band ($F_{hc}$) and full-band ($F_{tc}$) fluxes by propagating  errors already obtained for $MedE_t$, $F_h$, and $F_t$.  This propagation uses the absorption correction calibrations in Figure \ref{fig_redcorr_vs_mede_sim} and Table \ref{tbl_calibration3}.  As with $\log N_H$, statistical errors  on intrinsic source fluxes are estimated using the best spectral model, while  systematic errors are estimated using the ``second nearest neighbor'' models.

\section{Application to M17 X-ray Sources \label{m17_section}}

We now illustrate our analysis methods using the $Chandra$ ACIS-I X-ray data of PMS candidate stars in the M17 region. M17, one of the brightest HII regions in the sky, is ionized by OB stars in the rich young stellar cluster NGC~6618 which lies on the edge of a massive molecular cloud. The X-ray stellar population is composed of light to moderately-absorbed PMS stars of the central NGC~6618 cluster supplemented by heavily-absorbed stars embedded in or obscured by the surrounding cloud.  A study of the X-ray population from a 40~ks {\it Chandra} ACIS exposure of the region is described by \citet{Broos07};  886 faint X-ray sources were found in this short exposure.

A new $\sim 300$~ks $Chandra$ exposure of this field has been obtained, and the combined dataset yields $\sim 2000$ X-ray sources (Townsley et al., in preparation).  This dataset is chosen for its large sample and wide range of source properties: $5 \la NC_t \la 1000$ source counts and $1 \la MedE_t \la 5$~keV median energies. The $Chandra$ data have been analyzed following {\it ACIS Extract} procedures described in \citet{Broos02,Broos09}. Spectral fits have been performed with {\it XSPEC} on sources with net counts down to $\sim 10$~counts; best-fit models were found by the maximum likelihood method\footnote{The C-statistic was used to fit ungrouped spectra for more than 2000 M17 sources. The background spectra available for most sources contained $\sim 100$ counts.  Within XSPEC these moderate-quality backgrounds were modeled with a simple continuous piecewise-linear function with 10 approximately evenly-spaced vertices (see the discussion of the {\em cplinear} model in the AE manual, \url{http://www.astro.psu.edu/xray/docs/TARA/ae\_users\_guide.pdf}). In order to explore the range of alternative model fits the TBABS~$\times$~VAPEC model with thawed parameters was fit to each source five times, using a set of initial parameter values that explore an appropriate region of parameter space. Two additional TBABS~$\times$~VAPEC models representing ``standard'' stellar spectra (kT frozen at 0.86  and at 2.6 keV) were also fit. The best model derived from these seven fits was chosen via a combination of computer algorithm and human review. The full procedure requires days to be accomplished.} assuming a one-temperature TBABS $\times$ VAPEC model \citep{Wilms00,Smith01} with the elemental abundances fixed at the values typical for PMS or extremely active zero-age main-sequence stars from \citet{Gudel07}.  These are slightly different spectral models to those used in the COUP study, but the small change in spectral shape is likely to have negligible effect on the inferred absorption and flux estimates.
 
The M17 data thus give a large sample of PMS X-ray sources for which spectral and photometric properties are available.  The absorption and flux estimates derived with the parametric and nonparametric methods can thus be compared. 
  
\subsection{X-ray column densities towards M17 sources \label{nh_m17_subsection}}

\begin{figure}
\centering
\includegraphics[angle=0.,width=5.8in]{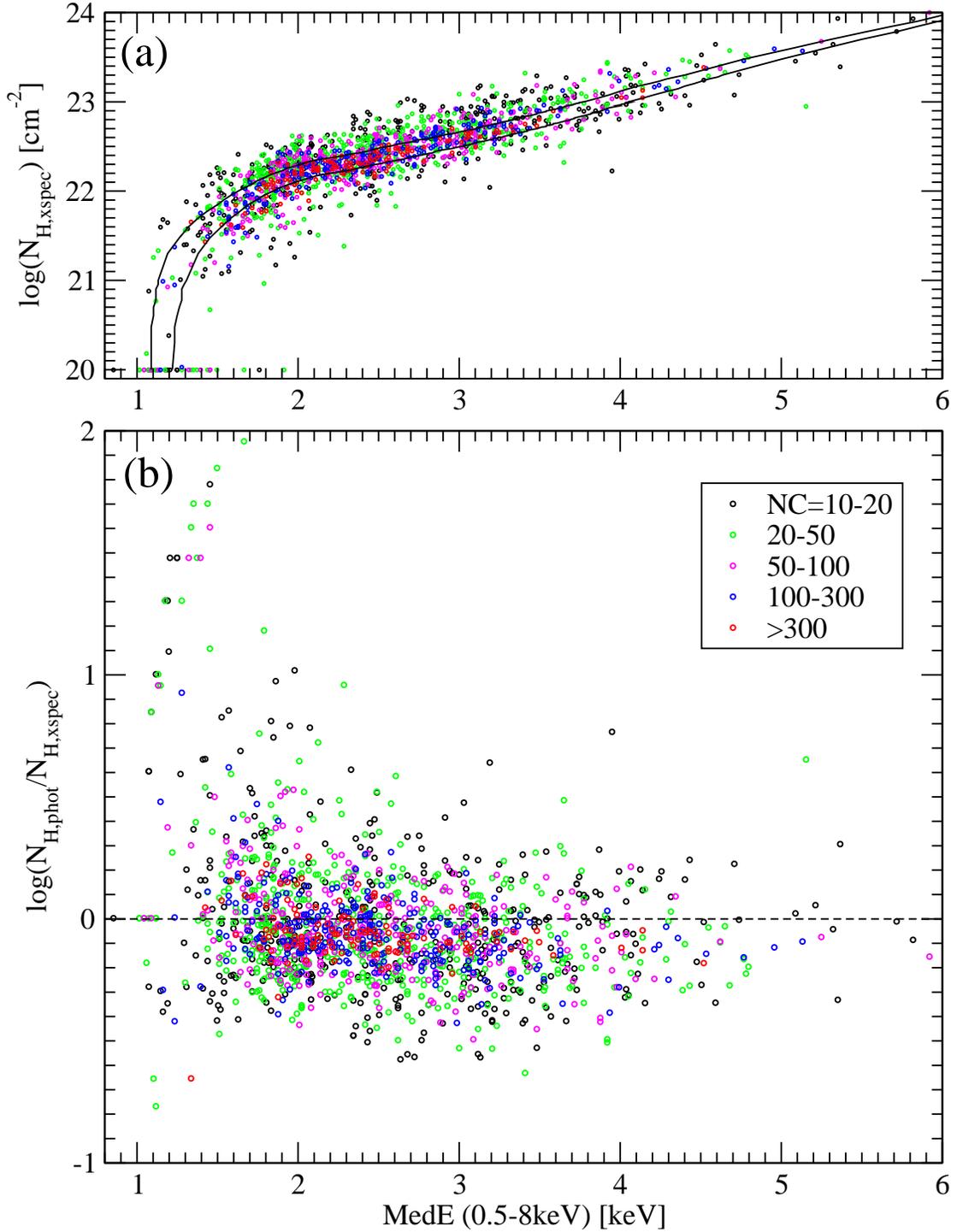}
\caption{(a) X-ray column density inferred from XSPEC spectral 1-temperature model fits of $>1600$ M17 sources plotted against X-ray median energy. Two solid curves are calibration lines corresponding to the simulated model families of $\log(L_{hc}) = 29$~ergs s$^{-1}$ (top) and $\log(L_{hc}) = 31$~ergs s$^{-1}$ (bottom) from Figure \ref{fig_nh_vs_mede_sim}. (b) Ratio of X-ray column density obtained from nonparametric estimation to that inferred from {\it XSPEC} spectral fits of the M17 data plotted against X-ray median energy. Net count strata are color-coded. \label{fig_nhratio_vs_mede_sim}}
\end{figure}

Figure \ref{fig_nhratio_vs_mede_sim} compares the nonparametric and parametric estimation of absorption for M17 sources. Panel (a) shows the $MedE_t$ obtained nonparametrically plotted against the column density $\log N_H$  obtained from {\it XSPEC}. The data accurately follow the relationship shown in Figure~\ref{fig_nh_vs_mede_sim} with increased scatter for fainter sources, as expected. Panel (b) shows the ratio of $N_H$ values estimated from $MedE_t$ as described in \S~\ref{mede_section} with those obtained from {\it XSPEC} fits, as a function of $MedE_t$. For the $MedE_t > 1.7$~keV range, the $\log N_H$ value inferred nonparametrically is systematically lower than that of the XSPEC fitting by 0.05~dex. This is a small effect, and we do not know if it represents a bias in our $MedE_t-\log N_H$ conversion, a bias in {\it XSPEC} fitting, or is due to the difference in spectral models used in the simulations and M17 spectral fitting. For sources with $MedE_t < 1.7$~keV, discrepancies between the nonparametric and parametric absorption estimates show much larger scatter and biases, up to a factor of ten. Neither method can accurately measure absorbing column density for faint soft sources, because $Chandra$ ACIS-I spectra are often insensitive to differences in low column density and there is an ambiguity in spectral model fits of faint soft sources.

\begin{figure}
\centering
\includegraphics[angle=0.,width=5.5in]{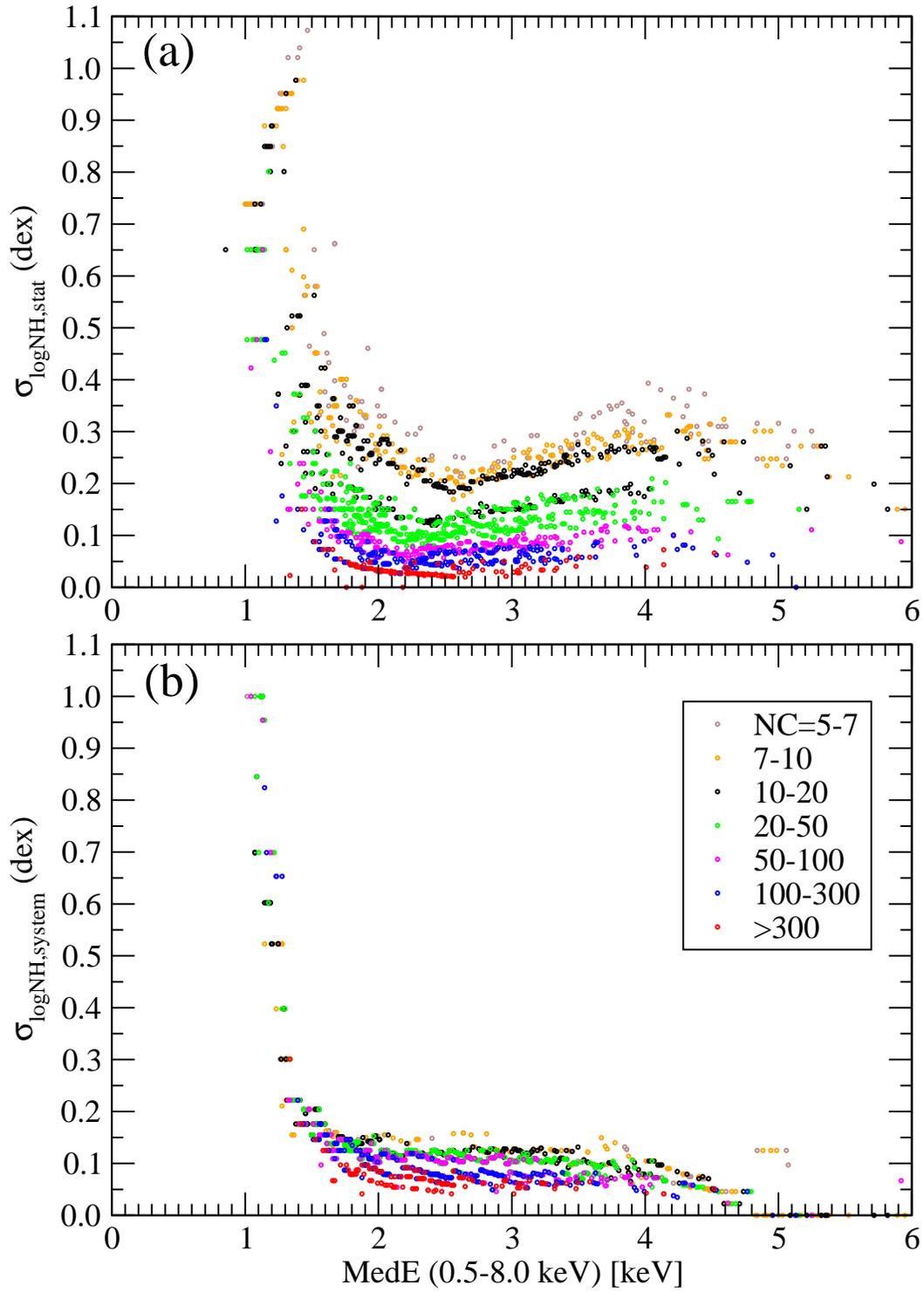}
\caption{Statistical (a) and systematic (b) errors on the column density $\log N_H$ inferred nonparametrically plotted against X-ray median energy. Net count strata are color-coded. Results are presented for $\sim 2000$ M17 sources. \label{fig_logNH_errors_vs_me}}
\end{figure}

We derive statistical and systematic errors on absorption column density following the procedure presented in \S \ref{method_error_section}. Figure \ref{fig_logNH_errors_vs_me} shows that $\log N_H$ values inferred from our method have increasing uncertainties for soft sources with $MedE_t < 1.7$~keV. Again, this arises from the weak dependency of apparent median energy on column density for softer sources with light obscuration (Figure \ref{fig_nh_vs_mede_sim}). We recommend that tabulations of nonparametric properties of PMS stars with $MedE_t < 1.7$~keV should report an upper limit of $\log(N_H) \la 22.0$~cm$^{-2}$ rather than specifying a very uncertain absorbing value.

For stars with $MedE_t > 1.7$~keV,  Figure \ref{fig_logNH_errors_vs_me}a shows that our method estimates $\log N_H$ with a statistical accuracy better than $\pm$0.3~dex, $\pm$0.2~dex, $\pm$0.1~dex for the $NC_t = 7-20$, $20-50$, $>50$ count strata, respectively.  Panel (b) shows that systematic uncertainties due to inadequate knowledge of the precise X-ray spectral model applicable to each source add an additional uncertainty less than $\pm 0.15$~dex for all stars, and less than $\pm 0.05$~dex for sources with $NC_t \ga 100$.   These errors are comparable to the scatter between photometric and {\it XSPEC} estimates of $\log N_H$ above 22.0~cm$^{-2}$ shown in Figure~\ref{fig_nhratio_vs_mede_sim}b.  In summary, the nonparametric procedure provides reliable estimates of $N_H$ absorption within a factor of $\sim 1.5$ for sources with $\ga 50$ counts, and within a factor of $\sim 2$ for sources with $\sim 10$ counts including both statistical and systematic spectral  model uncertainties.

\subsection{Apparent X-ray fluxes of M17 sources \label{flux_apparent_subsection}}

\begin{figure}
\centering
\includegraphics[angle=0.,width=6.0in]{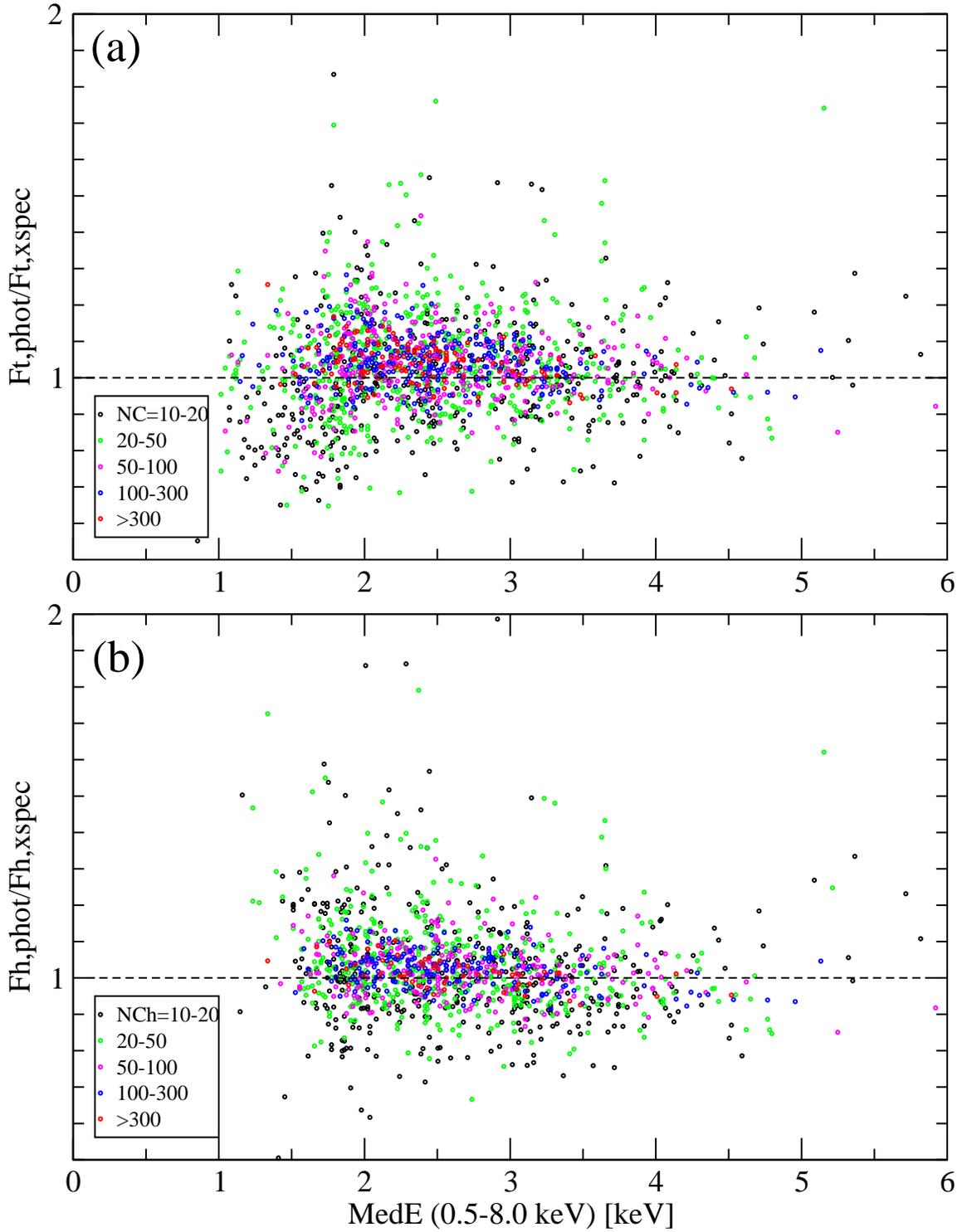}
\caption{Ratio of apparent (not corrected for absorption) de-biased photometric X-ray flux to that inferred from {\it XSPEC} spectral fitting of M17 data plotted against X-ray median energy. Results are presented for $\sim 1600$ M17 sources for the total band (panel a), and $\sim 1300$ M17 sources for the hard band (panel b). Net count strata are color-coded. Fluxes are in units of ergs s$^{-1}$ cm$^{-2}$. \label{fig_xspec_vs_phot_data}}
\end{figure}

Figure \ref{fig_xspec_vs_phot_data}a shows the ratio of the total-band apparent flux obtained using our nonparametric methods in \S~\ref{apparent_aefluxes_section} and the flux obtained from {\it XSPEC} fitting for M17 sources.  The biases shown in Figure \ref{fig_xspec_vs_phot_sim} have been removed from the nonparametric estimates.  The mean value is $1.04 \pm 0.10$ indicating that only a small ($<4$\%) systematic flux difference between the nonparametric and parametric methods remains, and individual source differences are mostly below $\pm 10$\%.  In Figure~\ref{fig_xspec_vs_phot_data}b, similar distributions for hard-band apparent flux ratios are shown, again exhibiting small ($\sim 2$\%) bias.  These small systematic differences between nonparametric  and {\it XSPEC} results may become important in the $>100-300$ counts regime, and parametric methods are preferred for strong sources.

\begin{figure}
\centering
\includegraphics[angle=0.,width=5.8in]{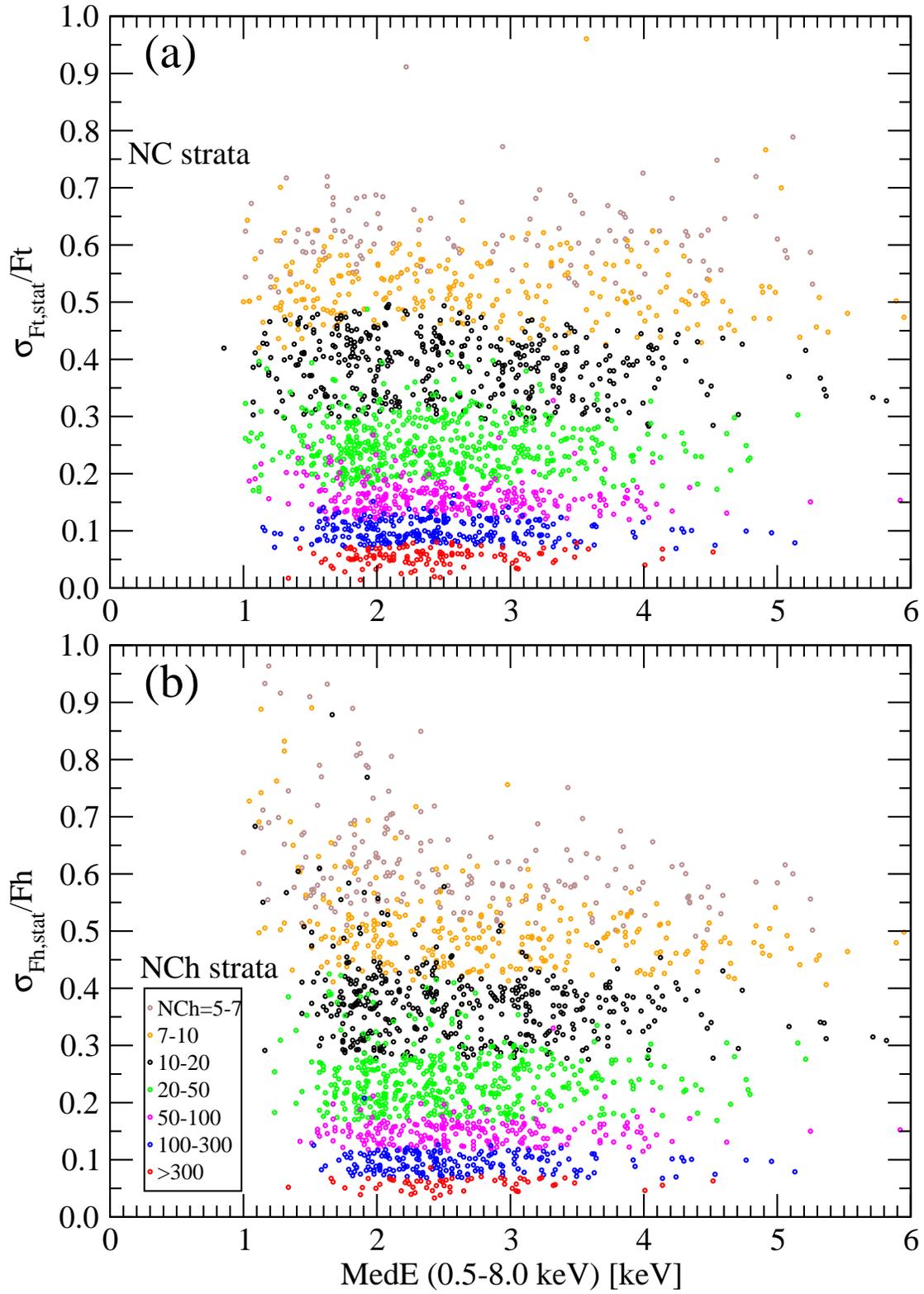}
\caption{Statistical errors on (a) the total band and (b) the hard band apparent X-ray fluxes inferred nonparametrically plotted against X-ray median energy for $\sim 2000$ and $\sim 1800$ M17 sources, respectively. Net count strata are color-coded. \label{fig_apparentflux_stat_errors_vs_me}}
\end{figure}

Following procedures outlined in \S~\ref{method_error_section}, we derive statistical errors on apparent fluxes and plot them in Figure~\ref{fig_apparentflux_stat_errors_vs_me}.  Panel $a$ shows that our method calculates apparent total-band source fluxes with statistical accuracy better than 60\%, 50\%, 30\%, 20\% for total-band net count strata of $7-10$, $10-20$, $20-50$, $>50$, respectively.  Panel $b$ shows similar errors for apparent hard-band source fluxes. These are slightly higher than Gaussian $\sqrt N$ errors, and reach $\sqrt{N}$ levels when $>100$ counts are present. 

\subsection{Absorption Corrected X-ray Fluxes of M17 Sources \label{corr_flux_subsection}}

\begin{figure}
\centering
\includegraphics[angle=0.,width=6.8in]{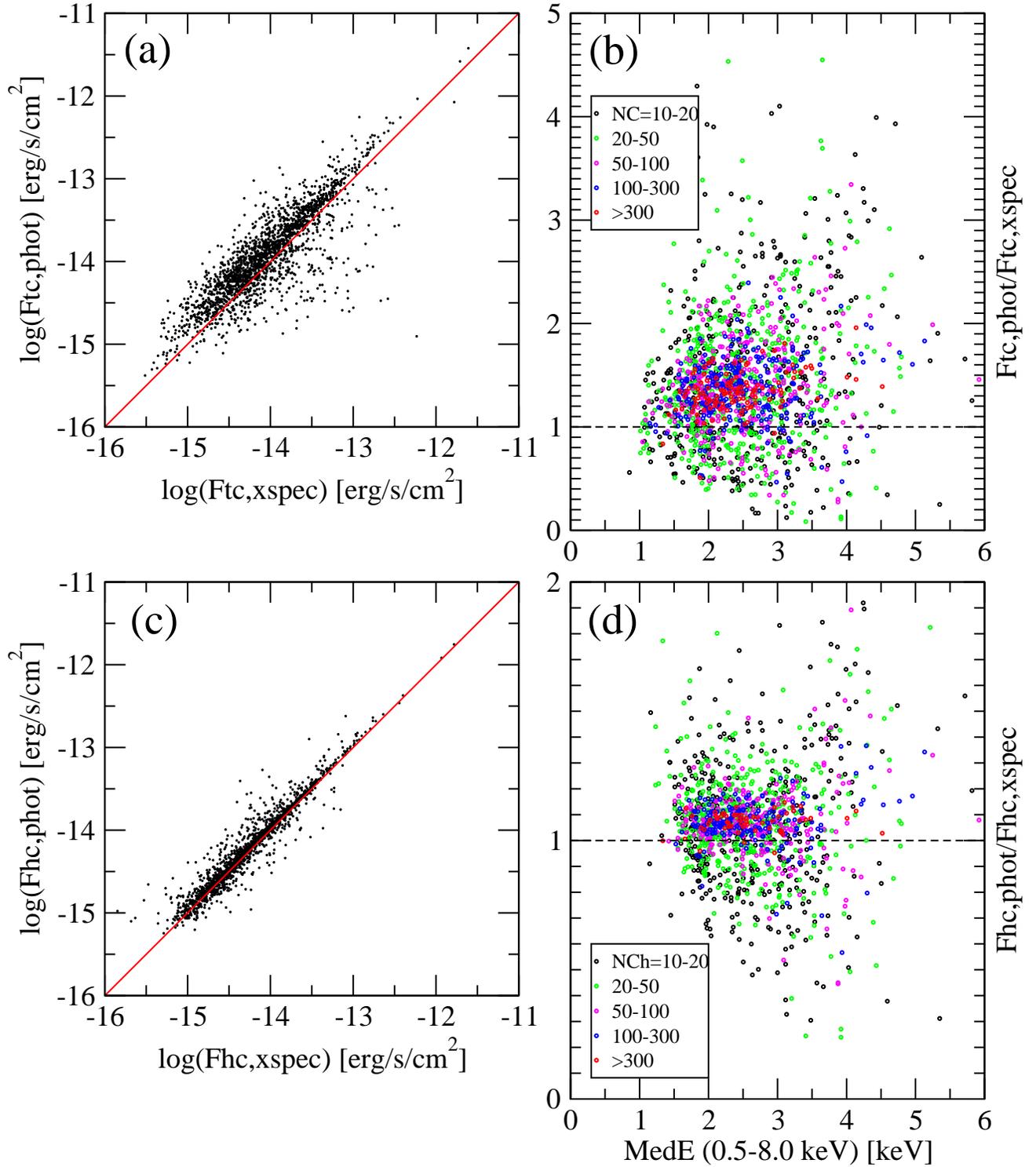}
\caption{Comparison of intrinsic fluxes (absorption-corrected) obtained from our nonparametric methods with those inferred from {\it XSPEC} spectral fits (panels a, c), and their ratios against X-ray median energy (panels b, d). Results are presented for $\sim 1600$ M17 sources for the total band (upper panels), and $\sim 1300$ M17 sources for the hard band (lower panels). Net count strata are color-coded. \label{fig_redcorr_simvsdata_vs_mede}}
\end{figure}

\begin{figure}
\centering
\includegraphics[angle=0.,width=6.0in]{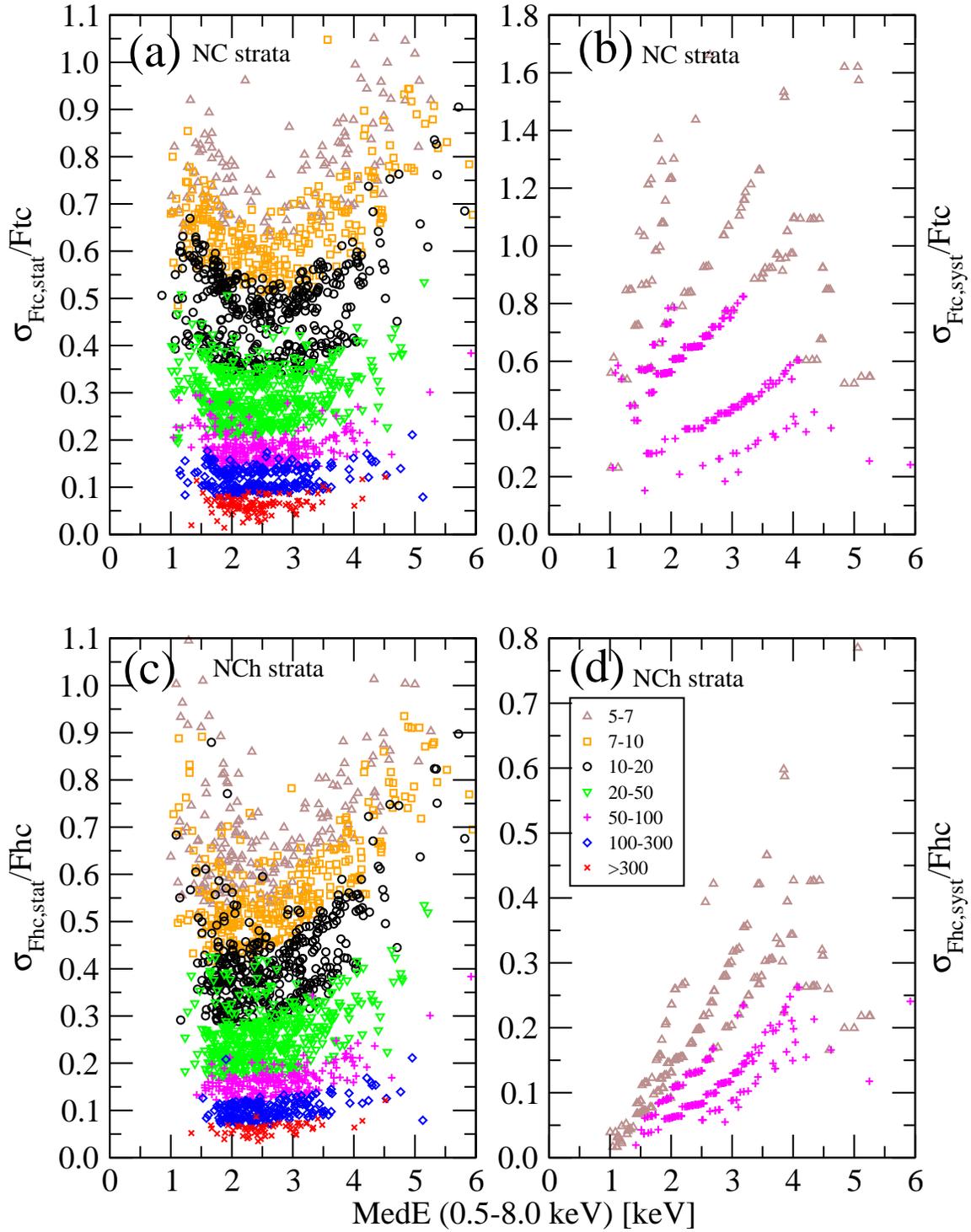}
\caption{Statistical (left panels) and systematic (right panels) errors of nonparametrically derived intrinsic fluxes corrected for absorption.  Results are presented for $\sim 2000$ M17 sources for the total band flux (upper panels) and for $\sim 1800$ M17 sources for the hard band (lower panels). Net count strata are color-coded. To avoid symbol clutter only $5-7$ and $50-100$ count strata are presented on panels (b) and (d). \label{fig_stat_syst_err_fluxcorr_vs_mede}}
\end{figure}

Figures \ref{fig_redcorr_simvsdata_vs_mede}-\ref{fig_stat_syst_err_fluxcorr_vs_mede} are similar to Figures~\ref{fig_xspec_vs_phot_data}-\ref{fig_apparentflux_stat_errors_vs_me} but for broad-band intrinsic fluxes of M17 sources corrected for absorption.  Recall that very large corrections have been applied to high-$MedE_t$ sources (Figure~\ref{fig_redcorr_vs_mede_sim}).  Here we see considerably greater bias and scatter compared to the apparent fluxes; note the expanded vertical scale in Figure~\ref{fig_redcorr_simvsdata_vs_mede}b.  The ratio of nonparametric to {\it XSPEC} intrinsic total-band fluxes is $1.45 \pm 0.74$ ($1.34 \pm 0.19$) for the $20-50$ ($>300$) total-band count strata, and the ratio of hard-band fluxes is $1.10 \pm 0.35$ ($1.07 \pm 0.04$) for the $20-50$ ($>300$) hard-band count strata. Thus, even for high-count rates, the intrinsic photometric total-band (hard-band) fluxes are systematically $\sim 35$\% ($\sim 7$\%) higher than {\it XSPEC} fluxes\footnote{Considering that the $\log N_H$ values inferred nonparametrically using two-temperature simulated models are systematically lower than that of the one-temperature XSPEC fitting (\S \ref{nh_m17_subsection}), it is clear that one-temperature models can not fully recover the soft intrinsic PMS X-ray emission subject to absorption.  This is further indicated by a comparison of 2-temperature model fits of the deep COUP dataset with 1-temperature model fits of an earlier $80$~ks $Chandra$ exposure \citep{Feigelson02}.  The COUP analysis gave $\sim 2$ times higher fluxes than the earlier analysis.  This result emphasizes the dangers of absorption correction in fluxes that include the soft band.}.

For any broad band flux that includes the soft $0.5-2$~keV band, the correction from apparent to intrinsic fluxes can be both very large (Figure~\ref{fig_redcorr_vs_mede_sim}a) and very uncertain (Figure~\ref{fig_stat_syst_err_fluxcorr_vs_mede}b), even for moderate absorption.  Weaker sources are particularly vulnerable.  We discourage use of the $F_{tc}$ and $L_{tc}$ quantities derived from photometry for scientific analysis. Although we have not examined it in detail, we suspect that $F_{tc}$ values from parametric {\it XSPEC}-type fitting will also be inaccurate and biased except for very strong sources.  

We derive statistical and systematic errors on intrinsic source fluxes following the procedure presented in \S \ref{method_error_section}. Figures~\ref{fig_stat_syst_err_fluxcorr_vs_mede}$a$,$b$ compare the inferred statistical and systematic errors on total-band intrinsic source flux. Here, systematic errors based on X-ray model uncertainty generally exceed statistical errors; again note the expanded vertical scale in panel $b$.  A maximal accuracy of $\pm 60\%$ (or $\pm$0.2~dex in log flux) is achieved only for sources with $>50$ total-band net counts.

In contrast to total-band absorption-corrected fluxes, the systematic errors of hard-band intrinsic fluxes do not exceed statistical errors for $<50$ hard-band count sources.  Here, systematic and statistical errors become comparable for brighter sources. An accuracy of $\sim 60\%$  for hard-band intrinsic fluxes can be achieved for sources with as few as $\sim 7-10$ hard-band net counts, unless very high absorptions ($MedE_t \ga 4$~keV) are present. We thus find, as expected from qualitative considerations, that hard-band absorption-corrected intrinsic fluxes are much more stable than intrinsic full-band fluxes. We thus encourage use of $F_{hc}$ when absorption-corrected fluxes and luminosities are sought using nonparametric methods.

\section{Applications to Other Classes of Faint X-ray Sources \label{other_classes_section}}

Recall from \S \ref{model_section} that a bright calibration sample of young stars in the Orion Nebula indicated a generic spectral model for this class of X-ray sources:  a two-temperature thermal plasma with the cooler temperature fixed and the hotter temperature scaled to X-ray luminosity (Figures \ref{fig_models} and \ref{fig_models_acc_strata}). This property of PMS stars allowed us to simulate X-ray models, and calibrate the derivation of astrophysical absorption and fluxes from the observed counts.  Other classes of X-ray sources will exhibit different intrinsic spectra shapes with somewhat different dependencies on observable quantities.
 
Spectral model families can be constructed for other classes of X-ray sources, extending nonparametric estimation methods beyond young star populations. Close binary star systems with accreting black holes typically show two spectral components, a thermal disk and a nonthermal powerlaw.  The relative strength of the components scales with the source luminosity and spectral hardness \citep{Remillard06}.  These relationships, based on bright Galactic X-ray binary systems, can be applied to X-ray binary systems in nearby spiral \citep[e.g.,][]{Griffiths00, Pietsch05, Stiele08} or elliptical \citep{Prestwich03, Sarazin03}  galaxies.  While the ultraluminous X-ray binaries in these galaxies may be bright enough for detailed parametric spectral modeling, the more typical source is well-adapted to our methods based on broad band counts and median energies.  Spectral model families might also be constructed for faint X-ray source populations in the Galactic Center region \citep{Muno03}.

The X-ray spectra of faint active galactic nuclei (AGN) are often parametrized by a power law with index $\Gamma$ subject to differing amounts of absorption, although more complex emission and absorption components may be present. At low redshift and luminosities, no statistical relationships are seen between $\Gamma$, luminosity, and absorption in the X-ray band \citep{Cappi06, Winter08}.  But at higher redshifts, quasar spectra appear to soften as luminosity increases \citep{Shemmer06,Saez08}. As there is substantial scatter in $\Gamma$ at both low and high redshifts, care must be taken to propagate uncertainties in spectral index through our procedures for estimating errors in the inferred broad-band fluxes and absorption.

\section{CONCLUSIONS \label{conclusions_section}}


In this work, we show that important properties of X-ray sources $-$ such as line-of-sight absorption, apparent broad-band fluxes, and intrinsic fluxes $-$  can be estimated with reasonable accuracy using the easily measured photometric quantities X-ray count rate and median energy.  The translation from observed to intrinsic quantities is achieved without employing often-uncertain and time consuming methods of non-linear parametric spectral modeling.  While parametric modeling is best for strong sources where spectral shape details are well-populated with photons, we believe nonparametric estimates are preferred for faint sources.  The concept of our procedures is similar to long-standing methods based on color-magnitude diagrams in optical and infrared astronomy. 

Specifically, we first demonstrate that the column density $\log N_H$ can be estimated directly from the total-band median energy $MedE_t$ with quantifiable uncertainties.  The apparent total-band ($0.5-8$~keV) and hard-band ($2-8$~keV) fluxes can be estimated from the source count numbers $NC_t$ and $NC_h$, and the absorption-corrected intrinsic fluxes $F_{hc}$ and $F_{tc}$ can then be derived. The methods are intended for sources with greater than $\sim 5-7$ counts but fewer than $100-300$ counts where parametric spectral fitting methods will be superior.  

We establish limitations to the method which are probably generally applicable for any reasonable spectral model family.  Nonparametric procedures provide poor measures of absorption for sources with $MedE_t < 1.7$~keV (or equivalently, $\log N_H < 22.0$ cm$^{-2}$).  Absorption corrections for total-band fluxes, $F_{tc}$, are often large and unreliable for sources with $< 50$ total-band counts.  However, absorption corrections for hard band fluxes, $F_{hc}$, are reliable down to $7-10$ hard-band counts\footnote{In the current implementation of the method the hard-band apparent flux (and thus hard-band net counts) as well as the full-band median energy are used as the primary input photometric quantities for derivation of intrinsic source properties (see Step 2 of our procedure described in \S \ref{translate_section}). As discussed in the paper, several net counts in the hard-band are needed in order to derive meaningful intrinsic properties. This level of hard-band signal may be difficult to detect from an intrinsically soft X-ray source. Researchers interested in non-parametric characterization of intrinsically soft weak X-ray sources might consider modifying the current implementation of the method to use the total-band apparent flux (instead of the hard-band flux) as a primary input photometric quantity.}.  $F_{hc}$ estimation using nonparametric techniques has small biases and uncertainties only somewhat larger than optimal $\sqrt{N}$ errors.

Quantitative results on biases, statistical errors, and systematic errors for each estimator are given in the text and figures. These detailed results are linked to assumptions of the intrinsic spectral models of the source populations, which we treat here to be absorbed two-temperature thermal plasmas associated with PMS stars in star-forming regions.  We calibrated the spectral models to the well-studied COUP sample, and successfully applied the methods to a new sample in the M17 star-forming region. 

The IDL implementation of our nonparametric method, {\it XPHOT.pro}, is provided.  Computing time needed to derive intrinsic properties for $\sim 2000$ X-ray sources is $<1$~minute. Our method and results can be directly used in statistical X-ray studies of young rich stellar populations; for example, for constructing column density maps and X-ray luminosity functions of young stellar clusters. Nonparametric results also can be used as initial and/or frozen parameters to use with the parametric {\it XSPEC} method for fine-tuning of spectral properties.

The procedures described here are developed in two specific contexts, but are more broadly applicable.  First, they use data products generated by the {\it ACIS Extract} data reduction package designed for the {\it Chandra} ACIS detector \citep{Broos02,Broos09}.  However, the methods are not restricted to this software package or detector; the necessary quantities such as median energy and count numbers can be readily obtained using CIAO, MIDAS, FTOOLS, or other packages.  The methods could be applied  to X-ray CCD observations with the {\it ASCA}, {\it XMM-Newton}, {\it Suzaku}, or other space-borne X-ray telescopes if the results are re-calibrated to each mirror-detector combination.  

Second, the analysis is conducted for specific two-temperature thermal plasma spectral models associated with PMS stars, and our quantitative results are restricted to this application.  However, similar analysis, likely resulting in similar results, can be made for other classes of sources, such as active galactic nuclei and X-ray binary star systems in nearby galaxies. Researchers extending these methods to such classes should explicitly define and simulate their own models to construct calibration tables to use with {\it XPHOT.pro}.

\acknowledgements We thank Michael Eracleous (Penn State) for helpful discussions, and the anonymous referee for helpful comments. This work is supported by the Chandra ACIS Team (G. Garmire, PI) through the SAO grant SV4-74018 and NASA Astrophysics Data Program grant NNX09AC74G (E. Feigelson, PI).

\clearpage

\begin{deluxetable}{cccc}
\centering \rotate \tabletypesize{\tiny} \tablewidth{0pt}
\tablecolumns{4}
\tablecaption{Template X-ray Spectral Models for COUP stars \label{tbl1}}
\tablehead{

\colhead{$\log(L_{hc})$} & \colhead{$kT_1$} & \colhead{$kT_2$} & \colhead{$EM_2/EM_1$}\\

(ergs~$s^{-1}$)& (keV) & (keV) &\\

(1)&(2)&(3)&(4)}

\startdata
27.0 & 0.8 & 0.8 & 0.2\\
27.5 & 0.8 & 1.1 & 0.4\\
28.0 & 0.8 & 1.5 & 0.6\\
28.5 & 0.8 & 1.8 & 1.0\\
29.0 & 0.8 & 2.2 & 1.3\\
29.5 & 0.8 & 2.5 & 1.9\\
30.0 & 0.8 & 2.8 & 2.5\\
30.5 & 0.8 & 3.2 & 2.8\\
31.0 & 0.8 & 3.5 & 3.2\\
31.5 & 0.8 & 3.9 & 3.2\\
32.0 & 0.8 & 4.2 & 3.2\\
\enddata

\tablecomments{Column 1: Hard-band intrinsic luminosity. Column 2: Temperature of the cool plasma component. Column 3: Temperature of the hot plasma component. Column 4: Ratio of emission measures between hot and cool components.}
\end{deluxetable}

\clearpage \clearpage

\begin{deluxetable}{cccc}
\centering \rotate \tabletypesize{\tiny} \tablewidth{0pt}
\tablecolumns{4}
\tablecaption{Simulated Data to Figure \ref{fig_xspec_vs_phot_sim}  \label{tbl_calibration1}}
\tablehead{

\colhead{$MedE_t$} & \colhead{$F_{phot}/F_{sim}$} & \colhead{$\log(L_{hc})$} & \colhead{Band}\\

(keV)& & (ergs~s$^{-1}$) &\\

(1)&(2)&(3)&(4)}

\startdata
0.971 & 1.049 & 27.0 & full\\
0.986 & 1.072 & 27.0 & full\\
1.000 & 1.104 & 27.0 & full\\
1.029 & 1.146 & 27.0 & full\\
1.059 & 1.187 & 27.0 & full\\
1.102 & 1.229 & 27.0 & full\\
1.146 & 1.266 & 27.0 & full\\
1.190 & 1.298 & 27.0 & full\\
1.234 & 1.323 & 27.0 & full\\
1.292 & 1.342 & 27.0 & full\\
\enddata

\tablecomments{Column 1: Apparent full-band median energy. Column 2: Ratio of the apparent X-ray flux of simulated data from {\it ACIS Extract} photometry to that inferred from flux integration of the input models evaluated on the energy bins set by the {\it Chandra} response matrix. Column 3: Hard-band intrinsic luminosity indicates corresponding simulated X-ray spectral model family from Table \ref{tbl1}. Column 4: Energy band, ``hard'' or ``full''. This table is available in its entirety in a machine-readable form in the online journal. A portion is shown here for guidance regarding its form and content.}
\end{deluxetable}

\clearpage \clearpage

\begin{deluxetable}{ccc}
\centering \rotate \tabletypesize{\tiny} \tablewidth{0pt}
\tablecolumns{3}
\tablecaption{Simulated Data to Figure \ref{fig_nh_vs_mede_sim}  \label{tbl_calibration2}}
\tablehead{

\colhead{$MedE_t$} & \colhead{$\log(N_H)$} & \colhead{$\log(L_{hc})$}\\

(keV)& (cm$^{-2}$) & (ergs~s$^{-1}$) \\

(1)&(2)&(3)}

\startdata
0.971 & 20.000 & 27.0\\
0.986 & 20.658 & 27.0\\
1.000 & 20.977 & 27.0\\
1.029 & 21.301 & 27.0\\
1.059 & 21.477 & 27.0\\
1.102 & 21.602 & 27.0\\
1.146 & 21.699 & 27.0\\
1.190 & 21.778 & 27.0\\
1.234 & 21.845 & 27.0\\
1.292 & 21.903 & 27.0\\
\enddata

\tablecomments{Column 1: Apparent full-band median energy. Column 2: X-ray column density. Column 3: Hard-band intrinsic luminosity indicates corresponding simulated X-ray spectral model family from Table \ref{tbl1}. This table is available in its entirety in a machine-readable form in the online journal. A portion is shown here for guidance regarding its form and content.}
\end{deluxetable}

\clearpage \clearpage

\begin{deluxetable}{cccc}
\centering \rotate \tabletypesize{\tiny} \tablewidth{0pt}
\tablecolumns{4}
\tablecaption{Simulated Data to Figure \ref{fig_redcorr_vs_mede_sim}  \label{tbl_calibration3}}
\tablehead{

\colhead{$MedE_t$} & \colhead{$\log (F_{c}/F)$} & \colhead{$\log(L_{hc})$} & \colhead{Band}\\

(keV)& & (ergs~s$^{-1}$) &\\

(1)&(2)&(3)&(4)}

\startdata
0.971 & 0.01311 & 27.0 & full\\
0.986 & 0.06390 & 27.0 & full\\
1.000 & 0.11212 & 27.0 & full\\
1.029 & 0.23303 & 27.0 & full\\
1.059 & 0.33018 & 27.0 & full\\
1.102 & 0.41713 & 27.0 & full\\
1.146 & 0.49527 & 27.0 & full\\
1.190 & 0.56576 & 27.0 & full\\
1.234 & 0.62951 & 27.0 & full\\
1.292 & 0.68732 & 27.0 & full\\
\enddata

\tablecomments{Column 1: Apparent full-band median energy. Column 2: Intrinsic to apparent flux ratio of simulated data. These are flux integrations of the input models (not re-fitted simulated spectra) evaluated on the energy bins set by the {\it Chandra} response matrix. Column 3: Hard-band intrinsic luminosity indicates corresponding simulated X-ray spectral model family from Table \ref{tbl1}. Column 4: Energy band, ``hard'' or ``full''. This table is available in its entirety in a machine-readable form in the online journal. A portion is shown here for guidance regarding its form and content.}
\end{deluxetable}

\clearpage \clearpage

\begin{deluxetable}{ccccc}
\centering \rotate \tabletypesize{\tiny} \tablewidth{0pt}
\tablecolumns{5}
\tablecaption{Simulated Data to Figure \ref{fig_mede_errors}  \label{tbl_calibration4}}
\tablehead{

\colhead{$MedE$} & \colhead{$\Delta MedE$} & \colhead{$\log(L_{hc})$} & \colhead{Band} & \colhead{$NC$}\\

(keV)& (keV)& (ergs~s$^{-1}$) & & (cnts)\\

(1)&(2)&(3)&(4)&(5)}

\startdata
2.431 & 0.2672 & 27.0 & hard &5\\
2.445 & 0.2716 & 27.0 & hard &5\\
2.460 & 0.2784 & 27.0 & hard &5\\
2.475 & 0.2868 & 27.0 & hard &5\\
2.489 & 0.2923 & 27.0 & hard &5\\
2.504 & 0.2981 & 27.0 & hard &5\\
2.518 & 0.3052 & 27.0 & hard &5\\
2.533 & 0.3131 & 27.0 & hard &5\\
2.548 & 0.3180 & 27.0 & hard &5\\
2.562 & 0.3212 & 27.0 & hard &5\\
\enddata

\tablecomments{Column 1: Apparent median energy in the corresponding energy band. Column 2: Statistical error on median energy. Column 3: Hard-band intrinsic luminosity indicates corresponding simulated X-ray spectral model family from Table \ref{tbl1}. Column 4: Energy band, ``hard'' or ``full''. Column 5: Net counts in the corresponding energy band. This table is available in its entirety in a machine-readable form in the online journal. A portion is shown here for guidance regarding its form and content.}
\end{deluxetable}

\clearpage \clearpage


\clearpage

\clearpage

\clearpage

\clearpage

\clearpage

\clearpage

\clearpage

\clearpage

\clearpage

\clearpage
 
\clearpage

\clearpage

\clearpage

\clearpage

\clearpage


\begin{thebibliography}

\bibitem[Anders \& Grevesse(1989)]{Anders89} Anders, E., \& Grevesse, N.\ 1989, \gca, 53, 197

\bibitem[Arnaud(1996)]{Arnaud96} Arnaud, K.\ A.\ 1996, in Data Analysis Software and Systems V, ed. G.\ H.\ Jacoby \& J.\ Barnes (San Francisco:ASP), 17

\bibitem[Bevington \& Robinson(1992)]{Bevington92} Bevington, P.~R., \& Robinson, D.~K.\ 1992, New York: McGraw-Hill, |c1992, 2nd ed. 


\bibitem[Broos et al.(2002)]{Broos02} Broos, P. S., Townsley, L. K., Getman,
K., \& Bauer, F. E.\ 2002, ACIS Extract, An ACIS Point Source Extraction
Package (University Park: The Pennsylvania State Univ.)
\url{http://www.astro.psu.edu/xray/docs/TARA/ae\_users\_guide.html}


\bibitem[Broos et al.(2007)]{Broos07} Broos, P.~S., Feigelson, E.~D., Townsley, L.~K., Getman, K.~V., Wang, J., Garmire, G.~P., Jiang, Z., \& Tsuboi, Y.\ 2007, \apjs, 169, 353

\bibitem[Broos et al.(2009)]{Broos09} Broos, P.~S., et al. in prep.

\bibitem[Cappi et al.(2006)]{Cappi06} Cappi, M., et al.\ 2006, \aap, 446, 459

\bibitem[Drake et al.(2006)]{Drake06} Drake, J.~J., Ratzlaff, P., Kashyap, V., Edgar, R., Izem, R., Jerius, D., Siemiginowska, A., \& Vikhlinin, A.\ 2006, \procspie, 6270, 49

\bibitem[Feigelson et al.(2002)]{Feigelson02} Feigelson, E.~D., Broos, P., Gaffney, J.~A., Garmire, G., Hillenbrand, L.~A., Pravdo, S.~H., Townsley, L., \& Tsuboi, Y.\ 2002, \apj, 574, 258

\bibitem[Feigelson et al.(2005)]{Feigelson05} Feigelson, E.~D., et al.\ 2005, \apjs, 160, 379 

\bibitem[Gagn{\'e} et al.(2004)]{Gagne04} Gagn{\'e}, M., Skinner, S.~L., \& Daniel, K.~J.\ 2004, \apj, 613, 393 

\bibitem[Getman et al.(2005)]{Getman05} Getman, K.~V., Flaccomio, E., Broos, P.~S. et al.\ 
2005, \apjs, 160, 319

\bibitem[Getman et al.(2008)]{Getman08} Getman, K.~V., Feigelson, E.~D., Broos, P.~S., Micela, G., \& Garmire, G.~P.\ 2008, \apj, 688, 418 

\bibitem[Griffiths et al.(2000)]{Griffiths00} Griffiths, R.~E., Ptak, A., Feigelson, E.~D., Garmire, G., Townsley, L., Brandt, W.~N., Sambruna, R., \& Bregman, J.~N.\ 2000, Science, 290, 1325 

\bibitem[G{\"u}del et al.(2007)]{Gudel07} G{\"u}del, M., et al.\ 2007, \aap, 468, 353 

\bibitem[Hong et al.(2004)]{Hong04} Hong, J., Schlegel, E.~M., \& Grindlay, J.~E.\ 2004, \apj, 614, 508 

\bibitem[Imanishi, Koyama, \& Tsuboi(2001)]{Imanishi01} Imanishi, K., Koyama, K., \& Tsuboi, Y.\ 2001, \apj, 557, 747

\bibitem[Maggio et al.(2007)]{Maggio07} Maggio, A., Flaccomio, E., Favata, F., Micela, G., Sciortino, S., Feigelson, E.~D., \& Getman, K.~V.\ 2007, \apj, 660, 1462

\bibitem[Mewe(1991)]{Mewe91} Mewe, R.\ 1991, \aapr, 3, 127

\bibitem[Morrison \& McCammon(1983)]{Morrison83} Morrison, R.~\& McCammon, D.\ 1983, \apj, 270, 119

\bibitem[Muno et al.(2003)]{Muno03} Muno, M.~P., et al.\ 2003, \apj, 589, 225

\bibitem[Pietsch et al.(2005)]{Pietsch05} Pietsch, W., Freyberg, M., \& Haberl, F.\ 2005, \aap, 434, 483 

\bibitem[Preibisch et al.(2005)]{Preibisch05} Preibisch, T., et al.\ 2005, \apjs, 160, 401

\bibitem[Press et al.(1992)]{Press92} Press, W.~H., Teukolsky, S.~A., Vetterling, W.~T., \& Flannery, B.~P. ``Numerical Recipes in C. The art of scientific computing.'' \ 1992, Cambridge: University Press, 2nd ed.  

\bibitem[Prestwich et al.(2003)]{Prestwich03} Prestwich, A.~H., Irwin, J.~A., Kilgard, R.~E., Krauss, M.~I., Zezas, A., Primini, F., Kaaret, P., \& Boroson, B.\ 2003, \apj, 595, 719 

\bibitem[Remillard \& McClintock(2006)]{Remillard06} Remillard, R.~A., \& McClintock, J.~E.\ 2006, \araa, 44, 49 

\bibitem[Saez et al.(2008)]{Saez08} Saez, C., Chartas, G., Brandt, W.~N., Lehmer, B.~D., Bauer, F.~E., Dai, X., \& Garmire, G.~P.\ 2008, \aj, 135, 1505 

\bibitem[Sarazin et al.(2003)]{Sarazin03} Sarazin, C.~L., Kundu, A., Irwin, J.~A., Sivakoff, G.~R., Blanton, E.~L., \& Randall, S.~W.\ 2003, \apj, 595, 743 

\bibitem[Shemmer et al.(2006)]{Shemmer06} Shemmer, O., Brandt, W.~N., Netzer, H., Maiolino, R., \& Kaspi, S.\ 2006, \apjl, 646, L29 

\bibitem[Smith et al.(2001)]{Smith01} Smith, R.~K., Brickhouse, N.~S., Liedahl, D.~A., \& Raymond, J.~C.\ 2001, \apjl, 556, L91 

\bibitem[Stiele et al.(2008)]{Stiele08} Stiele, H., Pietsch, W., Haberl, F., \& Freyberg, M.\ 2008, \aap, 480, 599 

\bibitem[Telleschi et al.(2007)]{Telleschi07} Telleschi, A., G{\"u}del, M., Briggs, K.~R., Audard, M., \& Palla, F.\ 2007, \aap, 468, 425 

\bibitem[Townsley et al.(2006)]{Townsley06} Townsley, L.~K., Broos, P.~S., Feigelson, E.~D., Garmire, G.~P., \& Getman, K.~V.\ 2006, \aj, 131, 2164 


\bibitem[Wilms et al.(2000)]{Wilms00} Wilms, J., Allen, A., \& McCray, R.\ 2000, \apj, 542, 914 

\bibitem[Winter et al.(2008)]{Winter08} Winter, L.~M., Mushotzky, R.~F., Tueller, J., \& Markwardt, C.\ 2008, \apj, 674, 686 


\end{thebibliography}
\end{document}